\documentclass[%
nofootinbib,
 amsmath,amssymb,
 aps,
 prd,
]{revtex4-2}

\usepackage{graphicx}% Include figure files
\usepackage{dcolumn}% Align table columns on decimal point
\usepackage{bm}% bold math
\usepackage{xcolor}
\usepackage{algorithm}
\usepackage{algpseudocode}	% Algorithms and pseudocodes
\usepackage{booktabs} % For professional looking tables
\usepackage{multirow}
\usepackage{longtable}
\usepackage{tabularx}

\usepackage[normalem]{ulem}

\newcommand{\best}{\textsc{CMB-BEST}}    % CMB-BEST
\newcommand{\vv}{\mathbf}			% Vectors
\newcommand{\bs}{\boldsymbol}		% Bold
\newcommand{\ells}{{\ell_1 \ell_2 \ell_3}}

\newcommand{\fNL}{f_\mathrm{NL}}
\newcommand{\kmin}{k_\mathrm{min}}
\newcommand{\kmax}{k_\mathrm{max}}

\newcolumntype{P}[1]{>{\centering\arraybackslash}p{#1}}     % Centered paragraph in a table

\begin{document}
%\printlength\textwidth

%\preprint{APS/123-QED}

\title{CMB bispectrum constraints on DHOST inflation}

\author{Wuhyun Sohn}
 \email{wuhyun@kasi.re.kr}
\affiliation{%
Korea Astronomy and Space Science Institute, Daejeon 34055, South Korea
}

\author{Andrei Lazanu}
 \email{andrei.lazanu@manchester.ac.uk}
\affiliation{%
Department of Physics and Astronomy, University of Manchester, Manchester, M13 9PL, UK
}
\author{Philippe Brax}
\email{philippe.brax@ipht.fr}
\affiliation{Institut de Physique Th\'eorique, Universit\'e Paris-Saclay,CEA, CNRS, F-91191 Gif-sur-Yvette Cedex, France.}

\author{James R. Fergusson}
\email{J.Fergusson@damtp.cam.ac.uk}
\affiliation{Centre for Theoretical Cosmology, Department of Applied Mathematics and Theoretical Physics, University of Cambridge, Cambridge CB3 0WA, United Kingdom.}

\date{\today}% It is always \today, today,
             %  but any date may be explicitly specified
\hspace{10pt}

\begin{abstract}

We present the first direct constraints on a Degenerate Higher Order Scalar Tensor (DHOST) inflation model using the  \textit{Planck} 2018 Cosmic Microwave Background (CMB) results on non-Gaussianities. 
We identify that the bispectrum consists of a fixed contribution following from the power spectrum and a linear combination of terms depending on five free parameters defining the cubic perturbations to the DHOST model. The former peaks in the squeezed limit, while the latter is maximised in the equilateral limit. We directly confront the model predictions to the CMB bispectrum statistics via the public code CMB-BEST and marginalize over the free parameters. {We explicitly show that there are viable DHOST inflationary models satisfying both power spectrum and bispectrum constraints from \textit{Planck}. However, {rather surprisingly}}, the constraints exclude certain models at the $6\sigma$-level even though they pass the conventional fudge factor tests. In this case and despite having a handful of free parameters, the model's large squeezed bispectrum cannot be cancelled out without introducing a large bispectrum in other limits which are strongly constrained by \textit{Planck}'s non-detection of primordial non-Gaussianity. We emphasize that first-order approximations such as fudge factors, albeit commonly used in the literature, may be misleading and provide weaker constraints. A proper analysis of the constraints from \textit{Planck} requires a more robust approach, such as the one provided by the CMB-BEST code. 

\end{abstract}

\maketitle

\section{Introduction}
The $\Lambda$CDM model has been successful in describing the properties and the evolution of the Universe, as confirmed by the \textit{Planck} satellite \cite{PlanckCollaboration2018Parameters}. This model is based on the assumption that the seeds of structures are primordial quantum fluctuations linked to a short inflationary epoch \cite{Guth1981inflation,Mukhanov:1981xt}.

The properties and the consequences of such an epoch have been extensively investigated and a large number of models have been proposed \cite{Hawking:1982cz, Kachru:2003sx, baumann_mcallister_2015,Martin:2013tda}. The anisotropies in the Cosmic Microwave Background (CMB) have been measured with exquisite accuracy by \textit{\textit{Planck}}, which has been able to constrain inflation at an unprecedented level \cite{PlanckCollaboration2018inflation}. The most stringent constraints come from measuring the power spectrum of primordial fluctuations, which is close to scale-invariant, but higher-order statistics, such as the three-point correlation function (bispectrum) have also been constrained. 

General relativity has been confirmed by numerous observations in the past century, but the recently confirmed late-time expansion of the Universe requires at least a cosmological constant. A large number of theories have been developed to explain this late-time cosmic acceleration, and a subset of these is represented by scalar-tensor theories. Degenerate Higher Order Scalar Tensor (DHOST) theories are the most general scalar-tensor theories involving one scalar field and which propagate one scalar degree of freedom \cite{Langlois:2015cwa, Achour:2016rkg, BenAchour:2016fzp, Crisostomi:2016czh, Crisostomi:2018bsp, Bombacigno:2021bpk} and no ghosts \cite{Ostrogradsky:1850fid}. They generalise Horndeski \cite{Horndeski1974} and beyond-Horndeski  \cite{Zumalacarregui:2013pma,Gleyzes:2014dya, Gleyzes:2014qga} theories. The similarity of the early and late-time evolution of the Universe makes similar types of theories viable for both epochs. Hence, although mostly used in the late Universe to explain its accelerated expansion, DHOST theories can also be applied to the inflationary epoch of the early Universe. Pure DHOST theories, however, are not compatible with experiments as they lead to a scale-invariant power spectrum with the spectral index $n_\mathrm{s}=1$, which is ruled out by \textit{Planck} at a $8\sigma$ level \cite{PlanckCollaboration2018Parameters}.

In \cite{Brax2021ps} models compatible with the current and future observational bounds at the power spectrum level have been presented, and in \cite{Brax2021bis} the models have been extended to the bispectrum level. A non-trivial spectral index for the primordial fluctuations can be obtained when the DHOST theory dictates the behaviour of the background cosmology and a small perturbation in the form of a potential is introduced. This potential breaks the shift-invariance of the DHOST models and can be seen to generate the correct behaviour at the background level. Explicitly we chose an axion potential as the leading shift-symmetry breaking potential. As the physics of the perturbations at the power spectrum level are mostly dependent on the mass of the scalar field and its quartic coupling, this is a fairly general approach which does not restrict the generality of the results. Indeed the quadratic and quartic corrections to the DHOST action are the leading corrections in an effective field theory sense for models with a $\mathbb{Z}_2$ symmetry.  In \cite{Brax2021bis}, it has been shown that the bispectrum constraints for the local, orthogonal and equilateral templates can be easily satisfied although the primordial non-Gaussianities (PNGs) remained intrinsically large. The purpose of the present work is to perform a \textit{full-shape} bispectrum analysis without using fudge factors and to investigate whether such models are viable given the latest CMB data.

The \textit{Planck} 2018 temperature and E-mode polarisation data have placed the most stringent bounds on PNG by itself so far \cite{PlanckCollaboration2018} and therefore is an ideal dataset for constraining PNG expected from DHOST inflation models. The \best\ formalism developed in the recent work of some of the authors \cite{Sohn2023cmbbest} and the accompanying public Python package \cite{Sohn2023cmbbestcode} are well suited for this purpose. In this work, we replace the analytic expressions computed in \cite{Brax2021bis} into CMB-BEST to extensively investigate the shape of the PNG predicted by DHOST models and \textit{Planck}'s constraints on them. We also compare to the analysis performed using `fudge factors', an approximation commonly used in the literature. We find that the fudge factors can be misleading and provide examples where the full analysis with CMB-BEST should be carried out and produce reliable results on the parameter space of the models.

This paper is organised as follows. In section \ref{sec:inflation} we briefly review the DHOST inflationary models, in section \ref{sec:method} we present the methods used to constrain the bispectrum with \textit{Planck}. The results and discussions are detailed in section \ref{sec:results}, and we conclude in section \ref{sec:conclusions}.

\section{DHOST inflation}
\label{sec:inflation}
 In the next paragraphs, we will briefly review the formalism used to build the DHOST model of inflation perturbed by an axionic potential. Considering only second-order interactions in the scalar field, the most general DHOST action can be expressed as
\begin{eqnarray}\label{action-DHOST}
S = \int d^4 x \sqrt{-g} \Big[ F_0(\phi,X) + F_1(\phi,X) \Box \phi + F_2(\phi,X) R 
+ \sum_{i=1}^5 A_i(\phi,X) L_i \Big] \,,
\end{eqnarray}
where $X=g^{\nu\eta}\phi_{\nu}\phi_{\eta}$,  $\phi_{\nu}\equiv\nabla_{\nu}\phi$,  the sign convention is $(-,+,+,+)$ and $L_i$ are the five possible Lagrangians quadratic in the field $\phi$ and $A_i(\phi,X)$ their corresponding amplitudes with
\begin{align}\label{DHOST-L2s}
L_1 &= \phi_{\nu\eta} \phi^{\nu\eta} , \hspace{1cm} L_2 = (\Box \phi)^2 ,
\hspace{1cm} L_3 = \Box\phi \, \phi_{\nu}\phi^{\nu\eta} \phi_{\eta} , 
\nonumber \\
L_4 &= \phi^{\nu} \phi_{\nu\eta} \phi^{\eta\lambda}\phi_{\lambda} , 
\hspace{1cm} L_5 = (\phi_{\nu}\phi^{\nu\eta}\phi_{\eta})^2 \,.
\end{align}

To keep the theory ghost-free, the functions $F_i$ and $A_i$ have to satisfy a certain set of degeneracy conditions \cite{Crisostomi:2017, Crisostomi:2019}. Imposing these conditions, and making the assumption that the functions only depend on $X$, i.e. the model possesses a shift symmetry, the action is simplified to
\begin{eqnarray}\label{action-dhost}
S_{\rm D} = \int d^4 x \sqrt{-g} \bigg[ F_0(X) + F_1(X) \Box\phi + F_2(X) R
+ \frac{6 F_{2,X}^2}{F_2} \phi^{\nu} \phi_{\nu\eta} \phi^{\eta\lambda}\phi_{\lambda} \bigg]\,.
\end{eqnarray}
This action cannot be used as a viable inflationary model, because it predicts a flat power spectrum \cite{Brax2021ps}, incompatible with observations. This problem can be rectified by introducing perturbations through a potential interaction term
\begin{equation}
    S_{\rm V} = -\int d^4 x \sqrt{-g}V(\phi)
\label{eq:interaction}
\end{equation}
where $V(\phi)$ is a sufficiently small interaction, which can be treated as a perturbation to the background cosmology driven by the DHOST action.  We choose a potential of the form $V(\phi)= \mu^4 (\cos\frac{\phi}{f}-1)$ whose origin could be a non-perturbative breaking of the shift symmetry $\phi\to \phi+c$ like in the case of axions \cite{Marsh2015}. 

By performing an expansion about $\phi=0$, for $\phi \ll f$ this action becomes
\begin{equation}
S_{\rm V} = \int d^4 x \sqrt{-g} \bigg[ - \frac{m_{\rm phys}^2}{2}  \phi^2 - \frac{\lambda_{\rm phys}}{4!}  \phi^4 \bigg] \,,    
\end{equation}
where $m^2_{\rm phys}<0$  and 
\begin{equation}
m^2_{\rm phys}=- \frac{\mu^4}{f^2},\ \lambda_{\rm phys}= \frac{\mu^4}{f^4} \, .
\end{equation}

The variables can be expressed in terms of tilded dimensionless coordinates as
\begin{equation}\label{coordinets-ch}
{\tilde t} \equiv \Lambda t \,, \hspace{1cm} {\tilde x}^i \equiv \Lambda x^i \,,
\end{equation}
\begin{equation}\label{dimensionless-couplings}
\phi \equiv M \, \varphi\,, \hspace{.5cm} X\equiv{M^2 \Lambda^2}{\mathrm x}\,, \hspace{.5cm} 
F_0 \equiv \Lambda^4 f_0 \,, \hspace{.5cm}
F_1 \equiv \frac{\Lambda^2}{M} f_1 \,, \hspace{.5cm} F_2 \equiv \Lambda^2 f_2 \,.
\end{equation}
The parameter $f_2={\cal O}(1)$ is chosen such that  $\Lambda\simeq m_{\rm Pl}$, where time and space are measured in Planck length. The scale $M$ gives the typical excursion scale of the scalar field \cite{Vafa:2005ui, Bedroya:2019snp}. This scale is not constrained by the CMB but one can place an upper bound
$ M\lesssim 10^{-6} m_{\rm Pl}$ using the distance conjecture of string theory \cite{Ooguri:2006in}

We treat the DHOST action and its perturbation as a low-energy effective theory below the strong coupling scale $\mu_c$
\begin{equation}
\mu_c= \sqrt{M\Lambda}
\label{eq:muc}
\end{equation}
with $\mu_c\ll m_{\rm Pl}$.
Similarly, the perturbation can be written at leading order as
\begin{equation}
S_{\rm V} = \int d^4 \tilde x \sqrt{-\tilde g} \bigg[ - \frac{m^2}{2}  \varphi^2 - \frac{\lambda}{4!}  \varphi^4 \bigg] \,   
\end{equation}
where $g_{\mu\nu}= \Lambda^2 \tilde g_{\mu\nu}$ and $\tilde g_{\mu\nu}$ is the dimensional metric { depending on $\tilde t$ and $\tilde x$}.
In terms of the  dimensionless parameters $m^2$ and $\lambda$ we have
\begin{equation}
    f= \frac{\sqrt{\vert m^2 \vert}}{\sqrt \lambda} M, \ \mu= \frac{\sqrt{\vert m^2 \vert}}{ \lambda^{1/4}} \Lambda  \, .
\end{equation}
The procedure determining the power spectrum for DHOST theories has been described in \cite{Brax2021ps}, based on \cite{Gorji:2020bfl}, where the standard formalism of inflationary perturbation theory has been used \cite{Peter:2013avv}. In particular, the action is expanded to second order in perturbations in the comoving gauge and then the field is quantised using a modified Mukhanov-Sasaki variable. This allows one to determine the power spectrum. 

The bispectrum of the model is calculated by employing the formalism 
of \cite{Maldacena2003, Chen2010review}: the action is expanded at third order in the comoving gauge, the Hamiltonian in the interaction picture is determined and the three-point correlation function is then calculated using the in-in formalism. The entire procedure is described in detail in Ref. \cite{Brax2021bis}.

In the case of the power spectrum, the parameters of the model can be fixed using the CMB constraints on the amplitude of the scalar perturbations, the scalar spectral index and its two derivatives, as well as by fixing the tensor-to-scalar ratio and the Hubble constant. The perturbation parameters $m^2$ and $\lambda$ are fixed as well by these constraints.

The parameters of the models appearing up to second order in the derivatives of the functions $f_i$ can be expressed in terms of the $\alpha$ coefficients (first derivatives) \cite{Bellini:2014fua, Gleyzes:2014rba, Langlois:2017mxy, Motohashi:2017gqb}
\begin{equation}
\alpha_H \equiv - {\rm x} \frac{f_{2,{\rm x}}}{f_2}\,, \hspace{1cm}
\alpha_B \equiv \frac{1}{2} \frac{\dot{\varphi}\,{\rm x}}{h_b} \frac{f_{1,{\rm x}}}{f_2} + \alpha_H \,, \hspace{1cm}
\alpha_K \equiv - \frac{{\rm x}}{6h_b^2}\frac{f_{0,{\rm x}}}{f_2} + \alpha_H + \alpha_B \,, \label{eqn:alpha_parameters}
\end{equation}
and the $\beta$ coefficients (second order)
\begin{eqnarray}
&&\beta_K \equiv - \frac{{\rm x}^2}{3} \frac{f_{0,{\rm x}{\rm x}} }{h_b^2 f_2} 
+ (1-\alpha_H) (1+3 \alpha_B) + \beta_B 
+ \frac{(1 + 6 \alpha_H - 3 \alpha_H^2) \alpha_K
- 2 ( 2 - 6 \alpha_H + 3 \alpha_K ) \beta_H}{1-3 \alpha_H} \,,
\nonumber \\
&& \beta_B \equiv  \dot{\varphi}\,{\rm x}^2 \frac{f_{1,{\rm x}{\rm x}}}{h_b f_2} \,,
\hspace{1cm}
\beta_H \equiv {\rm x}^2 \frac{f_{2,{\rm x}{\rm x}}}{ f_2}
\,.     \label{eqn:beta_parameters}
\end{eqnarray}
All the $\alpha$ parameters are fixed at the power spectrum level together with $\beta_K$. In the case of the bispectrum, there are three third-order derivatives of the functions $f_i$, in addition to $\beta_B$ and $\beta_H$ {that are not constrained at the power spectrum level.}  Notice that the normalisation parameter $f_1$ does not appear in the final result. An analysis of the bispectrum was performed in \cite{Brax2021bis}, by projecting the resulting bispectrum shapes on the local, equilateral and orthogonal shapes using a method described in \cite{Babich:2004gb}. This projection allowed us to derive constraints on the parameters of the model in a significantly simplified formalism, with only the equilateral configuration of the triangles being required. We showed that, although the parameters of the models can be tuned to make the bispectrum arbitrarily small for these specific templates, the overall amplitude remains large and therefore these models are potentially detectable. A full likelihood analysis is then required to investigate this possibility. This is the purpose of this work. In the following sections, we will describe the formalism for making this analysis.

For the numerical calculations, we first fix the values of the parameters according to the prescription taken from \cite{Brax2021ps} and compatible with the \textit{Planck} power spectrum constraints, i.e. $f_2=2.70$, $\alpha_B=1$, $\alpha_H=1.04$, $\beta_K=3.97343$,  $m^2 = -1.6 \times 10^{-23}$, $\lambda =  10^{-36}$ and  $h_{\mathrm{ds}}=3 \times 10^{-5}$. We will see that these values are essentially excluded by the analysis of the non-Gaussianities. Other values of the parameters such as   $f_2 = 6$, $\alpha_B = 1$  $\alpha_H = 1.04$, $\beta_K = 3.97343$, $m^2 = -10^{-23}$, $\lambda = 1.2 \times 10^{-36}$ and $h_{\rm ds} = 0.00001712$ do not suffer from the same problem and are not excluded. These two values are taken to illustrate the point that the fudge factor analysis can be misleading. A thorough discussion of the parameter space is left for future work.

In \cite{Brax2021bis}, there was an apparent discontinuity in the derivative of the bispectrum $B(k_1,k_2,k_3)$ at the surfaces $k_i=k_j$ with $i\neq j$. This was simply because the bispectrum was evaluated at the time of horizon exit defined as $\eta_f = -1 / (c_\mathrm{s} \mathrm{max}\{ k_1,k_2,k_3 \})$, which contains a non-smooth maximum function. We mitigate this unphysical discontinuity by setting $\eta_f = -1 / (c_\mathrm{s} \sqrt{ k_1^2 + k_2^2 + k_3^2})$ in this work.  

\section{Methodology}
\label{sec:method}
\subsection{CMB bispectrum likelihood}

In this section, we review how the CMB bispectrum can be used to test theoretical predictions on the primordial bispectrum. A detailed review can be found, for example, in \cite{Komatsu2010,Liguori2010,Fergusson2012,Sohn2023cmbbest}.

Given a set of spherical harmonic coefficients $a_{\ell m}$'s of the CMB anisotropies, an estimate for the angle-averaged CMB bispectrum can be constructed as
\begin{align}
    \hat{B}_\ells \equiv \sum_{m_j} \begin{pmatrix} \ell_1& \ell_2& \ell_3 \\ m_1& m_2& m_3 \end{pmatrix} \Bigl[  a_{\ell_1 m_1} a_{\ell_2 m_2} a_{\ell_3 m_3} - \left[ \langle a_{\ell_1 m_1} a_{\ell_2 m_2} \rangle a_{\ell_3 m_3} + (2~\mathrm{cyc.})  \right] \Bigr], \label{eqn:observed_bispectrum}
\end{align}
where the Wigner 3-j symbols enforce the angular momentum conservation and the bracket $\left< \, \cdot \, \right>$ denotes expected value over realisations. The theoretical bispectrum is the expected value of this estimate predicted by theory:
\begin{align}
    B^\mathrm{th}_\ells \equiv \frac{1}{f_\mathrm{sky}} \langle \hat{B}_\ells \rangle 
    = h_\ells b^\mathrm{th}_\ells, \label{eqn:theory_bispectrum}
\end{align}
for some geometric factor $h$ that also enforces the angular momentum conservation \footnote{This is defined as $h_{\ell_1 \ell_2 \ell_3} \equiv \sqrt{\frac{(2\ell_1+1)(2\ell_2+1)(2\ell_3+1)}{4\pi}} \begin{pmatrix}
    \ell_1 & \ell_2 & \ell_3 \\ 0 & 0 & 0
\end{pmatrix} $, where the Wigner 3-j symbol has been used on the right hand side.}. The sky fraction $f_\mathrm{sky}$, which equals $0.78$ for the \textit{Planck} temperature maps used in our analysis, corrects for the lack of power due to the incomplete sky coverage. The reduced bispectrum appearing on the right-hand side is related to the primordial bispectrum via CMB transfer functions $T_\ell(k)$:
\begin{align}
    b^\mathrm{th}_\ells &= \left( \frac{2}{\pi} \right)^3 \int dr dk_1 dk_2 dk_3 \; (r k_1 k_2 k_3)^2  B_\zeta^{\mathrm{th}}(k_1,k_2,k_3) \prod_{j=1}^{3} \left[ j_{\ell_j}(k_j r) T_{\ell_j} (k_j) \right],
\end{align}
where the curvature bispectrum at the end of inflation is defined as
\begin{align}
    \langle \zeta_{\vv{k}_1} \zeta_{\vv{k}_2} \zeta_{\vv{k}_3} \rangle &= (2\pi)^3 \delta^{(3)}(\vv{k}_1 + \vv{k}_2 + \vv{k}_3) B_\zeta^\mathrm{th}(k_1,k_2,k_3).
\end{align}

Assuming that the CMB is at most weakly non-Gaussian, the CMB bispectrum likelihood of a given theory can be written as
\begin{align}
    L (\mathbf{B}^\mathrm{th} | \hat{\mathbf{B}}) \propto \mathrm{exp} \left[ -\frac{1}{2} \sum_{\ell_1 \le \ell_2 \le \ell_3} \frac{ \Delta_\ells }{ 6 f_\mathrm{sky}\; C_{\ell_1} C_{\ell_2} C_{\ell_3}} \left( \hat{B}_\ells - f_\mathrm{sky} B^\mathrm{th}_\ells \right)^2 \right].  \label{eqn:bispectrum_likelihood_formal}
\end{align}
The symmetry factor $\Delta_\ells=6,3,1$ for 3, 2, 1 distinct values of $\ell$'s, respectively, and vanishes when $(\ell_1,\ell_2,\ell_3)$ do not form three sides of a triangle.

\vspace{10pt}

Now, suppose that our theoretical bispectrum is a linear function of parameters $\bs{\theta}$. The primordial and late-time bispectra can then be written as a sum over the individual templates:
\begin{align}
    B_\zeta^\mathrm{th}(k_1,k_2,k_3;\bs{\theta}) &= \sum_{i=1}^{N_\mathrm{pars}} \theta_i B_\zeta^{(i)}(k_1,k_2,k_3), \\
    B^\mathrm{th}_\ells (\bs{\theta}) &= \sum_{i=1}^{N_\mathrm{pars}} \theta_i B^{(i)}_\ells.
\end{align}
For example, the \textit{Planck} analyses \cite{PlanckCollaboration2018} include a constraint for which $\bs{\theta}=(\fNL^\mathrm{equil}, \fNL^\mathrm{ortho})$ with the two corresponding bispectrum templates.
The likelihood \eqref{eqn:bispectrum_likelihood_formal} then becomes a function of $\bs{\theta}$;
\begin{align}
    L (\mathbf{B}^\mathrm{th}(\bs{\theta}) | \hat{\mathbf{B}}) \propto \mathrm{exp} \left[ \bs{\theta}^T \vv{s} - \frac{1}{2} \bs{\theta}^T F \bs{\theta} \right], \label{eqn:bispectrum_likelihood_simplified}
\end{align}
where
\begin{align}
    s_i &\equiv \sum_{\ell_1 \le \ell_2 \le \ell_3} \frac{ \Delta_\ells }{ 6\; C_{\ell_1} C_{\ell_2} C_{\ell_3}} B^{(i)}_\ells (\hat{B}_\ells - f_\mathrm{sky} B^{(0)}_\ells ),     \label{eqn:estimator_signal} \\
    F_{ij} &\equiv \sum_{\ell_1 \le \ell_2 \le \ell_3} \frac{ f_\mathrm{sky} \Delta_\ells }{ 6\; C_{\ell_1} C_{\ell_2} C_{\ell_3}} B^{(i)}_\ells B^{(j)}_\ells. \label{eqn:estimator_fisher_matrix} \\
\end{align}
The maximum likelihood estimator (MLE) for $\bs{\theta}$ is the global maximum of this likelihood function, given by
\begin{align}
    \hat{\bs{\theta}}^\mathrm{MLE} = F^{-1} \vv{s}. %\\
\end{align}

The Fisher matrix $F$ measures the amount of information contained in the data for constraining $\bs{\theta}$. In principle, the MLE is optimal and the expected covariance matrix is given by
\begin{align}
    \mathrm{Cov}[\hat{\bs{\theta}}^\mathrm{MLE}] = F^{-1}.
\end{align}
Note that the marginalized error for $\hat{\theta}_i^\mathrm{MLE}$ is given by $(F^{-1})_{ii}$ (no summation implied), as opposed to $(F_{ii})^{-1}$ expected for an independent analysis with $N_\mathrm{pars}=1$ and $\theta=\theta_i$. In our analysis, we utilise 160 end-to-end simulations with Gaussian initial conditions (FFP10 simulations, \cite{PlanckCollaboration2015simulations,PlanckCollaboration2018hfi}) to estimate the estimation error and account for realistic noise properties. 

We follow the \best\ formalism \cite{Sohn2023cmbbest} in this work. The theoretical \textit{primordial} bispectrum is decomposed in terms of separable bases as 
\begin{align}
	(k_1 k_2 k_3)^2 B^\mathrm{(i)}_\zeta (k_1, k_2, k_3) = \sum_{n \leftrightarrow (p_1, p_2, p_3)} \alpha^{(i)}_n \; q_{p_1}(k_1) q_{p_2}(k_2) q_{p_3}(k_3), \label{eqn:basis_expansion}
\end{align}
for $i=0,1,\cdots,N_\mathrm{pars}$ and basis functions $q_p(k)$. We choose the Legendre polynomials basis with $p_\mathrm{max}=30$ for an accurate expansion, although $p_\mathrm{max}=10$ has been found to yield constraints that are consistent to a sub percent level. The quantities \eqref{eqn:estimator_signal} and \eqref{eqn:estimator_fisher_matrix}  are given in terms of the coefficients $\alpha^{(i)}_n$ by

\begin{align}
    s_i &= \frac{1}{6} \left( \bs{\alpha}^{(i)} \right)^{T} \bs{\beta}, \\
    F_{ij} &= \frac{f_\mathrm{sky}}{6} \left( \bs{\alpha}^{(i)} \right)^{T} \Gamma \bs{\alpha}^{(j)},
\end{align}
where the matrix $\Gamma$ and the array $\bs{\beta}$ are precomputed for the Legendre polynomial basis  $\{ q_p (k) \}$ \footnote{To be precise, the basis is given by $q_0(k)=k^{n_\mathrm{s}-2}$ and $q_p(k)=P_{p-1} \left( -1+\frac{2(k-\kmin)}{\kmax-\kmin} \right)$ for $p=1,\cdots,p_\mathrm{max}$, where $P_n(x)$ is the Legendre Polynomial of order $n$. More details can be found in \cite{Sohn2023cmbbest}.} and provided together with the public code \cite{Sohn2023cmbbestcode}.

To evaluate how much our theory improves the fit to the data over the null hypothesis of vanishing bispectrum, we may compute the log Bayes factor defined as
\begin{align}
    \Delta \ln L &\equiv \ln \left( \frac{ L(\vv{B}^\mathrm{th} = \vv{B}^\mathrm{th}(\bs{\theta}) \ |\  \hat{\vv{B}}) } { L(\vv{B}^\mathrm{th}=\vv{0}\ |\ \hat{\vv{B}}) } \right)    \\
    &= \bs{\theta}^T \vv{s} - \frac{1}{2} \bs{\theta}^T F \bs{\theta}.   \label{eqn:log_likelihood_def}
\end{align}

\subsection{DHOST bispectrum}

Analytic expressions for the primordial bispectrum arising from DHOST inflation are given in \cite{Brax2021bis}. We write the DHOST bispectrum as
\begin{align}
    B^\mathrm{DHOST}(k_1,k_2,k_3;\; \boldsymbol{\theta}) = B^\mathrm{base}(k_1,k_2,k_3) + \sum_{i=1}^{5} \theta_i B^{(i)} (k_1, k_2, k_3),     \label{eqn:dhost_bispectrum_base}
\end{align}
where $\boldsymbol{\theta}=(f_{\rm 0,xxx}, f_{\rm 1,xxx}, f_{\rm2,xxx}, \beta_H, \beta_B)^\mathrm{T}$ is a vector containing five DHOST parameters. The base term captures all contributions that are independent of $\bs{\theta}$. Note that other parameters appearing in DHOST inflation are fixed at the power spectrum level to match the \textit{Planck} CMB power spectrum constraints.

\subsubsection{Fudge factors approach}

In the previous work \cite{Brax2021bis}, CMB constraints on the model were placed via `fudge factors'. The fudge factors approach, first introduced in \cite{Babich:2004gb} and somewhat commonly used in the literature including \cite{GarciaSaenz2018fudge,Koshelev2023fudge,Mirbabayi2023fudge}, provides rough constraints on a bispectrum shape through the standard constraints on the local, equilateral and orthogonal templates.

Suppose that a given bispectrum shape can be expressed as a linear combination of the standard templates as
\begin{align}
    S(k_1,k_2,k_3)  \approx \sum_{t \in \{ \mathrm{local,equil,ortho}\} } \mathcal{F}^{(t)} S^{(t)} (k_1,k_2,k_3),    \label{eqn:shape_function_as_templates}
\end{align}
for some weights $F^t$ along the standard local, equilateral and orthogonal templates. Here, we have defined the dimensionless shape function as $S(k_1,k_2,k_3)\equiv (k_1 k_2 k_3)^2 B(k_1, k_2, k_3)$. The inner product between shape functions is defined as a weighted integral \cite{Babich:2004gb}
\begin{align}
    S_1 \cdot S_2 \equiv \int dk_1 dk_2 dk_3 \, w(k_1,k_2,k_3) \,  S_1 (k_1,k_2,k_3) \,  S_2 (k_1,k_2,k_3), \label{eqn:shape_function_inner_product}
\end{align}
for example with $w(k_1,k_2,k_3)=(k_1+k_2+k_3)^{-1}$.

In this approach, the standard shapes are assumed to be orthogonal with respect to this inner product. The values of $\mathcal{F}^{(t)}$ can then be written as the fudge factors given by
\begin{align}
    \mathcal{F}^{(t)} = \frac{ S \cdot S^{(t)} }{ S^{(t)} \cdot S^{(t)} }.
\end{align}
These factors represent an orthogonal projection of the shape function to the function space spanned by the standard templates. \textit{Planck}'s CMB data places bounds on the amplitudes $F^t$ appearing in the shape function \eqref{eqn:shape_function_as_templates} through the standard templates \cite{PlanckCollaboration2018}:
\begin{align}
    \mathcal{F}^\mathrm{local} \approx \fNL^\mathrm{local} = -0.9 \pm 5.1, \quad
    \mathcal{F}^\mathrm{equil} \approx \fNL^\mathrm{equil} = -26 \pm 47, \quad
    \mathcal{F}^\mathrm{ortho} \approx \fNL^\mathrm{ortho} = -38 \pm 24,      \label{eqn:fudge_factor_constraints}
\end{align}
where the errors are given at $68\%$ confidence level \footnote{Note that the $\fNL^\mathrm{local}$ constraint here is from an independent constraint instead of the joint constraint marginalized over $\fNL^\mathrm{equil}$ and $\fNL^\mathrm{ortho}$, which is another layer of approximation in the fudge factor analysis.}. Therefore, computing the fudge factors provides a quick way of testing the consistency of a given bispectrum shape with the CMB bispectrum constraints. Alternatively, finding a set of model parameters with the fudge factors $\mathcal{F}^{(t)}$ equal to $1$, for example, may show that the model can be consistent. The bispectrum amplitude at the equilateral limit $k_1=k_2=k_3=k_*$ is also sometimes used to place approximate bounds. The pivot scale is $k_*=0.05\,\mathrm{Mpc}^{-1}$.

For the DHOST bispectrum given in \eqref{eqn:dhost_bispectrum_base}, the fudge factors can be written as
\begin{align}
    \mathcal{F}^{(t)} = \mathcal{F}^{\mathrm{base},(t)} + \sum_{i=1}^{5} \theta_i \mathcal{F}^{(i),(t)},  \label{eqn:dhost_fudge_factors}
\end{align}
where the fudge factor $\mathcal{F}^{(i),(t)}$ is defined between shape $S^{(i)}(k_1,k_2,k_3)=(k_1 k_2 k_3)^2 B^{(i)} (k_1,k_2,k_3)$ and template $t$. Note that the linear equations \eqref{eqn:dhost_fudge_factors} can almost always be solved for arbitrary values of the fudge factors, since there are five free parameters $\theta_i$ for three equations.
This fails only when some shapes are linearly dependent exactly and the equations \eqref{eqn:dhost_fudge_factors} are incompatible. Note that the equilateral fudge factor constraint is sometimes loosely enforced by setting $\fNL(k_*,k_*,k_*)\approx 0$.

Even if the approximation of \eqref{eqn:shape_function_as_templates} is valid, having small fudge factors that satisfy \eqref{eqn:fudge_factor_constraints} is a \textit{necessary but not sufficient} condition for being consistent with the CMB observational data; the model can have small fudge factors while being strongly disfavored by the data. The full shape information of the primordial bispectrum is required for sufficiency. We will discuss this again in the result section.

\subsubsection{Utilising the full bispectrum shape}

In this work, on the other hand, we make use of the full bispectrum shape and compare it directly with the CMB without the fudge factor approximations. The analytic formulae for \eqref{eqn:dhost_bispectrum_base} was first converted from a \texttt{Mathematica} expression to a \texttt{C} code and then wrapped into \texttt{Python} using the \texttt{Cython} library \cite{Behnel2010cython}.

\begin{figure*}[htbp!]
	\centering    
	\includegraphics[]{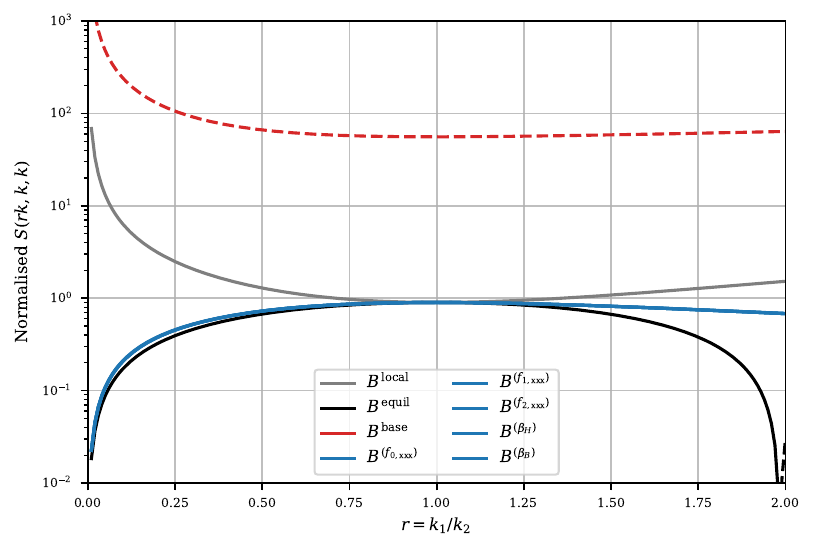}
	\caption{Bispectrum shape functions of the DHOST bispectrum components, together with the standard local and equilateral templates for comparison. The shape function is $S(k_1,k_2,k_3) \equiv (k_1 k_2 k_3)^2 B(k_1,k_2,k_3)$. The limit where $k_1=rk_2=rk_3$ is plotted against $r$ while keeping the overall scale $K=k_1+k_2+k_3$ fixed; $K=3k_*=0.15\mathrm{Mpc}^{-1}$. The points $r=0, 1, 2$ correspond to the squeezed, equilateral, and flattened (folded) limits, respectively. The DHOST shape functions are normalised so that $S(k_*,k_*,k_*)=9/10$ except for the `base' (red) contribution. The `base' (red) shape is maximised in the squeezed limit, while the other five shapes (blue, indistinguishable here) are maximised in the equilateral limit.}	
	\label{fig:bispectrum_shape_plot}
\end{figure*}

We first look at the shape dependence of $B^\mathrm{base}$ and $B^{(i)}$s. Figure \ref{fig:bispectrum_shape_plot} shows a one-dimensional slice of the shape functions corresponding to $B^\mathrm{base}$ and $B^{(i)}$, together with two standard templates, local and equilateral. The shape function values on a constant scale slice ($K=k_1+k_2+k_3=\mathrm{const}$) are plotted as a function of $r=k_1/k_2$, from the squeezed limit ($r\rightarrow 0$), through the equilateral limit ($r=1$), and ending at the flattened limit ($r=2$). Note that the shape functions have been rescaled to have the sample value at the pivot point: $k_*^6 B(k_*,k_*,k_*)=9/10$.

We find that $B^\mathrm{base}$ has a shape similar to the local template and is maximised at the squeezed limit ($k_1\rightarrow 0,\ k_2=k_3$). On the other hand, the normalised shapes of $B^{(1)},\cdots,B^{(5)}$, which correspond to the parameters $f_{\rm 0,xxx}, f_{\rm 1,xxx}, f_{\rm2,xxx}, \beta_H, \beta_B$, are maximised at the equilateral configuration ($k_1=k_2=k_3$). The shapes are nearly identical and cannot be distinguished from the plot. Note that they do not vanish at the flattened limit $k_1=k_2+k_3$, unlike the equilateral template.

\section{Results and discussion}
\label{sec:results}
\subsection{Fisher correlation analysis}
{In this part, we analyse a model whose parameters are the ones chosen in Ref. \cite{Brax2021ps} and used in the previous section when plotting the different shapes of the bispectrum.}
Constraining the DHOST parameters $\boldsymbol{\theta}$ is analogous to fitting the observed CMB bispectra with respect to six bispectrum templates $B^{(i)}$, except with an offset induced by the base term $\vv{B}^\mathrm{base}$. In terms of the late-time bispectra, this is equivalent to $\hat{B}_\ells \rightarrow \hat{B}_\ells - f_\mathrm{sky} B^\mathrm{base}_\ells$.

\begin{figure*}[htbp!]
	\centering    
	\includegraphics[width=0.7\textwidth]{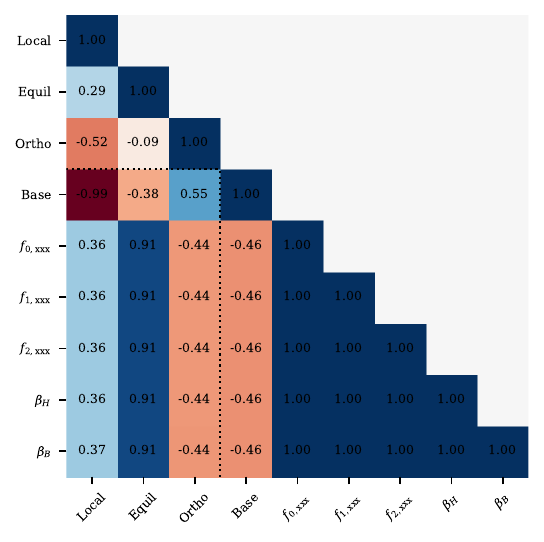}
	\caption{Correlation (`cosine') between the DHOST bispectrum components and the standard templates obtained from the Fisher matrix of the CMB bispectrum estimator. The `base' contribution is highly correlated with the local template, while the rest of the DHOST bispectrum components correlate with the equilateral template and between themselves. }	
	\label{fig:correlation}
\end{figure*}

First, we study the correlation between $B^\mathrm{base}$, $B^{(i)}$s and the standard templates to gain insight into their shapes and how well they can be distinguished via CMB bispectrum analyses. From the Fisher matrix $F_{ij}$ given by \eqref{eqn:estimator_fisher_matrix}, we compute the correlation between shapes by $C_{ij} = F_{ij}/\sqrt{F_{ii} F_{jj}}$ (no summation implied). The result is shown in Figure \ref{fig:correlation}.

The correlation values can be obtained from the Fisher matrix of the CMB bispectrum estimator. They indicate how similar the shapes are after the projection into the late-time CMB bispectra. Two shapes with a correlation close to 1 (or $-1$) probe nearly identical signatures in the CMB bispectrum, so it is difficult to distinguish the two using the CMB bispectrum estimator.

$B^\mathrm{base}$ correlates strongly with the local template, while the rest ($B^{(1)},\cdots,B^{(5)}$) correlate almost perfectly with each other, mostly with the equilateral template and partially with the orthogonal template. These are consistent with their shape dependence shown in Figure \ref{fig:bispectrum_shape_plot}.

Note that the strong correlations in Figure \ref{fig:correlation} have been rounded up to 1.00. Correlations between $B^{(1)}, B^{(2)}, B^{(3)}$, corresponding to $f_{\rm 0,xxx}, f_{\rm 1,xxx}, f_{\rm2,xxx}$, respectively, are of order $1-O(10^{-15})$. Correlations with any of the three and the ones for $\beta_H$ and $\beta_B$ are of order $1-O(10^{-9})$ to $1-O(10^{-6})$.

\subsection{Main constraints}

Since the bispectrum shapes for the parameters $\theta_1,\cdots,\theta_5$ are approximately identical, we combine them into a single shape parametrised with one degree of freedom. We introduce two new variables: $v_1$ and $v_2$ where
\begin{align}
    v_2 = \sum_{i=1}^{5} c_i \theta_i,    \label{eqn:v_definitions}
\end{align}
so that the DHOST bispectrum can be approximated as 
\begin{align}
    B^\mathrm{DHOST}(k_1,k_2,k_3) \approx v_1 B^\mathrm{base}(k_1,k_2,k_3) + v_2 \left(\frac{1}{5} \sum_{i=1}^{5} c_i^{-1} B^{(i)}(k_1,k_2,k_3) \right),
\end{align}
where $v_1=1$ corresponds to our model. The constants $c_i$ normalise the shapes as they appear in Figure \ref{fig:bispectrum_shape_plot}: $c_i^{-1}k_*^6 B^{(i)}(k_*,k_*,k_*)=9/10$ at $k_*=0.05\mathrm{Mpc}^{-1}$.

\begin{figure*}[htbp!]
	\centering
        \includegraphics[width=0.65\textwidth]{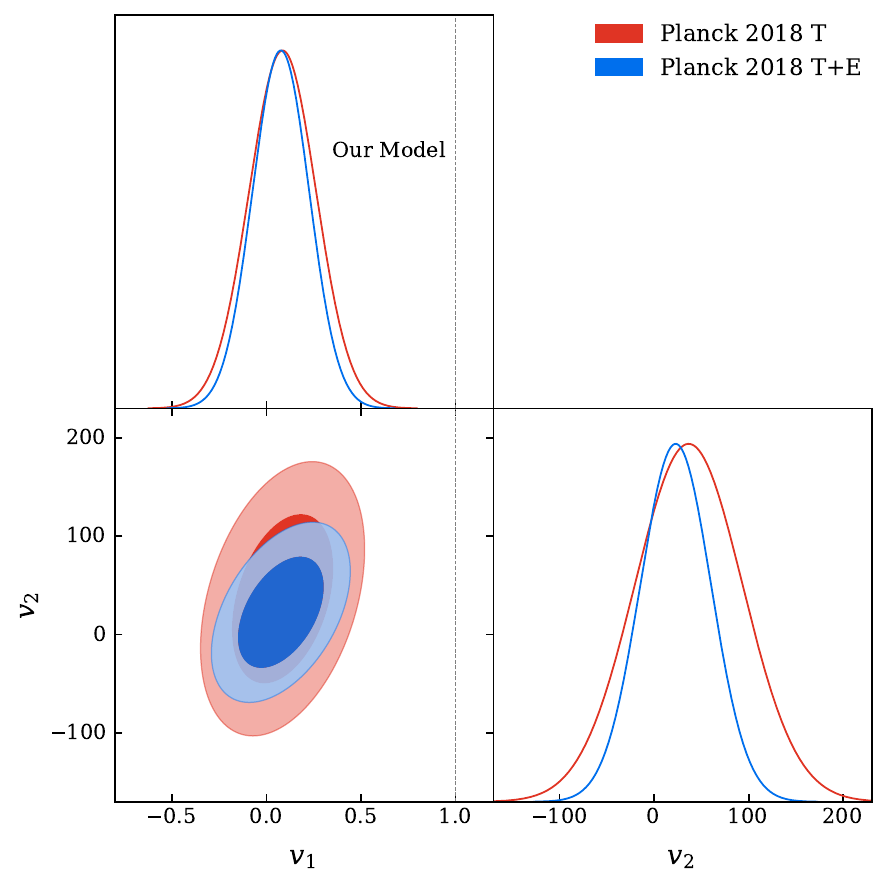}
	\caption{CMB bispectrum constraints on the two effective parameters $v_1$ and $v_2$ of the DHOST inflation model, where $v_1=1$ corresponds to our model. We use \textit{Planck} 2018 data via CMB-BEST \cite{Sohn2023cmbbest,Sohn2023cmbbestcode}, with (blue) and without (red) the E-mode polarisation data. We find that the model prediction $v_1=1$ lies at the $\sim 6.0\sigma$ away from the maximum in the marginalized constraints and therefore is disfavored by the data.}
	\label{fig:triangle_plot}
\end{figure*}

\begin{figure*}[htbp!]
	\centering
        \includegraphics[width=0.65\textwidth]{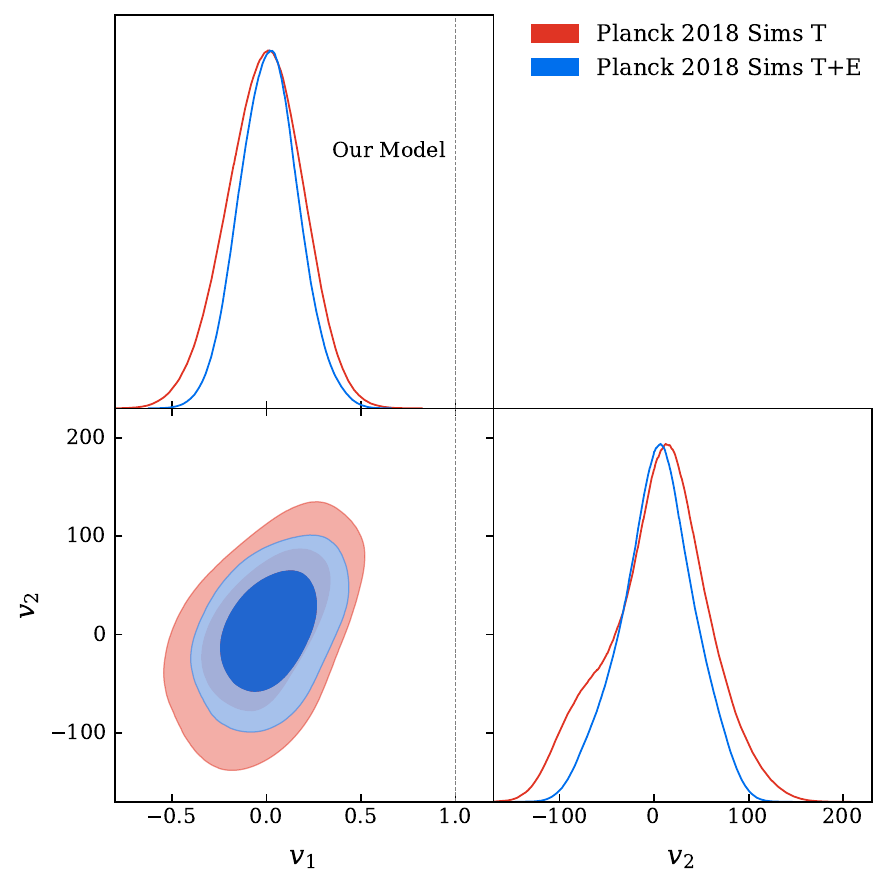}
	\caption{The distribution of the best-fit values of two effective parameters $v_1$ and $v_2$ in the DHOST inflation model, for 160 \textit{Planck} end-to-end simulations with Gaussian initial conditions, with (blue) and without (red) the E-mode polarisation data. Our model corresponds to $v_1=1$, which is unlikely in a Gaussian universe as shown in the plot. The contours are consistent with zero bispectrum ($v_1=v_2=0$), and the irregular shape of the contour is likely due to the limited number (160) of simulations used.}
	\label{fig:triangle_plot_samples}
\end{figure*}

Using the form above, we obtain the marginal constraints on $v_1$ and $v_2$ from \textit{Planck} 2018 data via \best\ \cite{Sohn2023cmbbest,Sohn2023cmbbestcode}. The results are summarised in the triangle plot shown in Figure \ref{fig:triangle_plot}. The Fisher error bars are used for the constraints shown as the CMB bispectrum estimator is shown to be nearly optimal. The results for \textit{Planck} end-to-end simulations with Gaussian initial conditions are shown in Figure \ref{fig:triangle_plot_samples} to follow.

First, we note that the constraints with and without the inclusion of E-mode polarisation data are statistically consistent with each other and with vanishing primordial non-Gaussianity ($v_1=v_2=0$). Polarisation data improves the constraining power by a factor of 1.2 and 1.6 for $v_1$ and $v_2$, respectively.

The constraints, however, are far away from our model prediction of $v_1=1$. The marginalized constraints for $v_!$ is $0.08 \pm 0.15$ and $0.09 \pm 0.19$ for T and T+E analyses, for which the point $v_1=1$ lies $6.0\sigma$ and $4.7\sigma$ away from the best-fit estimate, respectively. Our model with $v_1=1$ is therefore disfavored by the data.

In terms of the log Bayes factor defined in \eqref{eqn:log_likelihood_def}, the values for T+E and T-only analyses are $\Delta\ln L = -19.15$ and $-12.06$, respectively. The negativity of the Bayes factors means that the baseline of zero bispectrum is favoured over our model in both cases, despite having a set of extra free parameters.

The reason why our DHOST model has little support from the CMB bispectrum data is as follows. As seen in Figure \ref{fig:bispectrum_shape_plot}, our model has a large baseline contribution $B^\mathrm{base}$, about two orders of magnitude larger than the standard local template. This is large enough to be ruled out by the current non-detection of local PNG: $\fNL^\mathrm{local}=-0.9\pm 5.1$ \cite{PlanckCollaboration2018}. While the theory parameters $\bs{\theta}$ can be varied to cancel some of these large squeezed-limit contributions, they inevitably produce a large equilateral bispectrum ($\fNL^\mathrm{equil}\sim 5000$) to do so. Equilateral PNG is also constrained as $\fNL^\mathrm{equil}=-26\pm 47$, so this is not feasible either. In short, the current bounds on the local and equilateral PNGs rule our model out.

We note that our DHOST inflation model would also be ruled out in a universe with vanishing PNG. In Figure \ref{fig:triangle_plot_samples}, we show the best-fit values of $v_1$ and $v_2$ for 160 end-to-end simulations provided by \textit{Planck} \cite{PlanckCollaboration2018hfi} with Gaussian initial conditions. The distributions show that it is unlikely to find $v_1$ in a Gaussian-seeded universe. Furthermore, the distribution is narrow enough that it is not possible to shift the distribution and be consistent with both $v_1=0$ (vanishing PNG) and $v_1=1$ (this model). The data is consistent with $v_1=0$. We emphasize that it is the non-detection of PNG, rather than any PNG signal in the data, that provides us with enough constraining power to exclude this model.

\subsection{Limitation of fudge factor analyses}

\begin{figure*}[htbp!]
	\centering
        \includegraphics{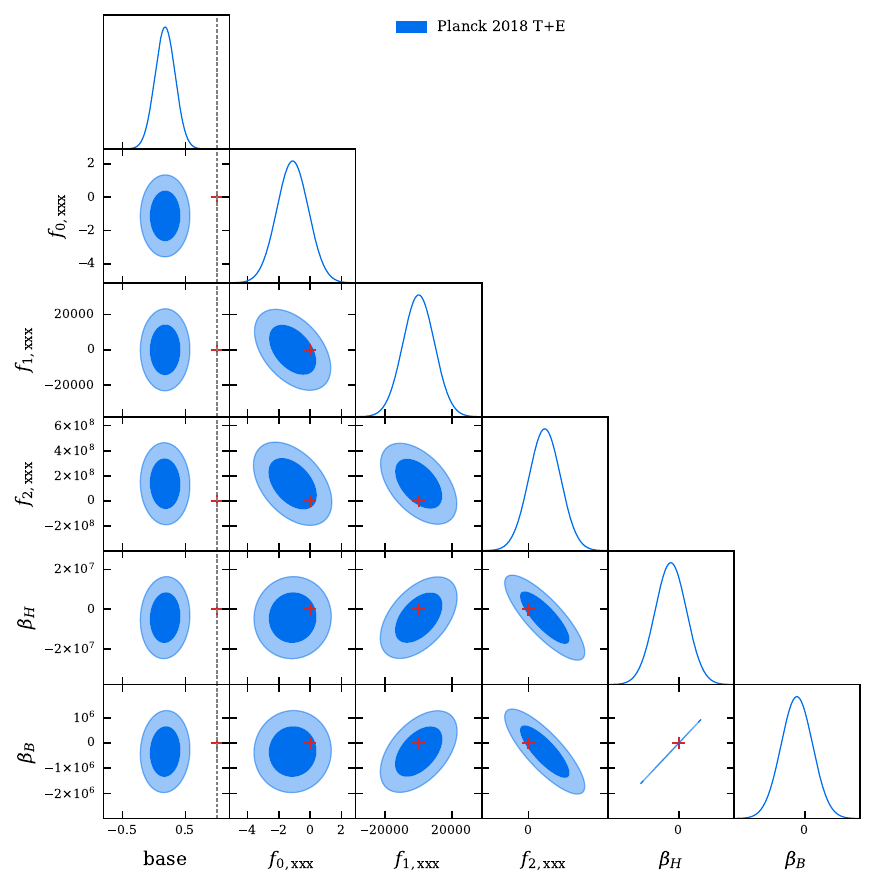}
	\caption{Joint constraints on all five DHOST parameters from \textit{Planck} CMB bispectrum, together with the `base' contribution parameterised as $v_1$ in Figure \ref{fig:triangle_plot}. The grey dashed line indicates that the model predicts a fixed amount for this base contribution. Plotted as red crosses are the parameter set $\bs{\theta}_\mathrm{fudge}$ that fix all three fudge factors identically to zero. While $\bs{\theta}_\mathrm{fudge}$ lies within the allowed range of the model parameters, the likelihood for the bispectrum is low due to the large base contribution.}
	\label{fig:triangle_plot_fudge_factors}
\end{figure*}

The results presented appear to contradict the fudge factor analysis; the three fudge factors \eqref{eqn:fudge_factor_constraints} can be fixed to be arbitrarily small using the 5 free parameters in the model. We find a set of parameters $\bs{\theta}_\mathrm{fudge}$ that satisfy $F^\mathrm{local}=F^\mathrm{equil}=F^\mathrm{ortho}=0$. This choice of model parameters is plotted with the full marginalized constraints in Figure \ref{fig:triangle_plot_fudge_factors}.

The marginalized constraints are again from CMB-BEST based on the \textit{Planck} 2018 temperature and polarisation data and are obtained while varying all five parameters of the $\bs{\theta}$. Due to the similarity in the bispectrum shapes for the parameters as seen in Figure \ref{fig:correlation}, the Fisher matrix is close to being degenerate and has many small eigenvalues. This results in large eigenvalues of the covariance matrix $\mathrm{Cov}[ \hat{\bs{\theta}^\mathrm{MLE}} ] = F^{-1}$ and hence provides bloated constraints; independent shape analyses on individual $\bs{\theta}$ provide orders of magnitude smaller bounds.

While the parameter choices from fudge factors analysis are consistent with the constraints on $\bs{\theta}$, the base contribution $v_1=1$ is still excluded by the data, meaning that the model is still disfavored at $6\sigma$ level. This shows that having small fudge factors does \textit{not} guarantee that the given bispectrum shape is consistent with the data. It is possible tune the model parameters so that the fudge factors are arbitrarily small, while the bispectrum amplitude is large enough to be ruled out by the data.

\subsection{Alternative model}

The model given in previous sections predicts a primordial bispectrum that is too large to be consistent with the CMB observations. In this model, we had fixed some of the DHOST inflation model parameters other than $\bs{\theta}$ to be consistent with the original works of \cite{Brax2021ps,Brax2021bis}. To be specific, $\alpha_H$, $\alpha_B$, $\alpha_K$ and $\beta_K$ given in \eqref{eqn:alpha_parameters} and \eqref{eqn:beta_parameters} were fixed to match the primordial scalar amplitude $A_\mathrm{s}$ and spectral index $n_\mathrm{s}$ from the \textit{Planck} CMB constraints. Note that there is some freedom in this choice as there are more variables than constraints.

While we have not written the dependence on the $\alpha$ variables and $\beta_K$ explicitly in the analytical form for the DHOST bispectrum given in \eqref{eqn:dhost_bispectrum_base}, they enter implicitly through the $B^\mathrm{base}(k_1,k_2,k_3)$ term.  Ideally, one could vary all possible parameters that appear in both the primordial power spectrum and bispectrum simultaneously and utilise both the CMB angular power spectrum and bispectrum statistics to constrain them. This is rather challenging due to the complexity of the analytic equations involved and the computational challenges in the bispectrum analyses. We leave this task as future work and simply exhibit one alternative set of parameters here. In this work, we showcase this alternative set of parameters that also satisfy the power spectrum constraints and are obtained by trial and error. The parameters are set to be $\alpha_H = 1.04$, $\alpha_B = 1$, $\beta_K = 3.97343$, $h_{\rm ds} = 0.00001712$, $f_2 = 6$, $\lambda = 1.2 \times 10^{-36}$, and $m^2 = 10^{-23}$.

\begin{figure*}[htbp!]
	\centering	\includegraphics[width=0.65\textwidth]{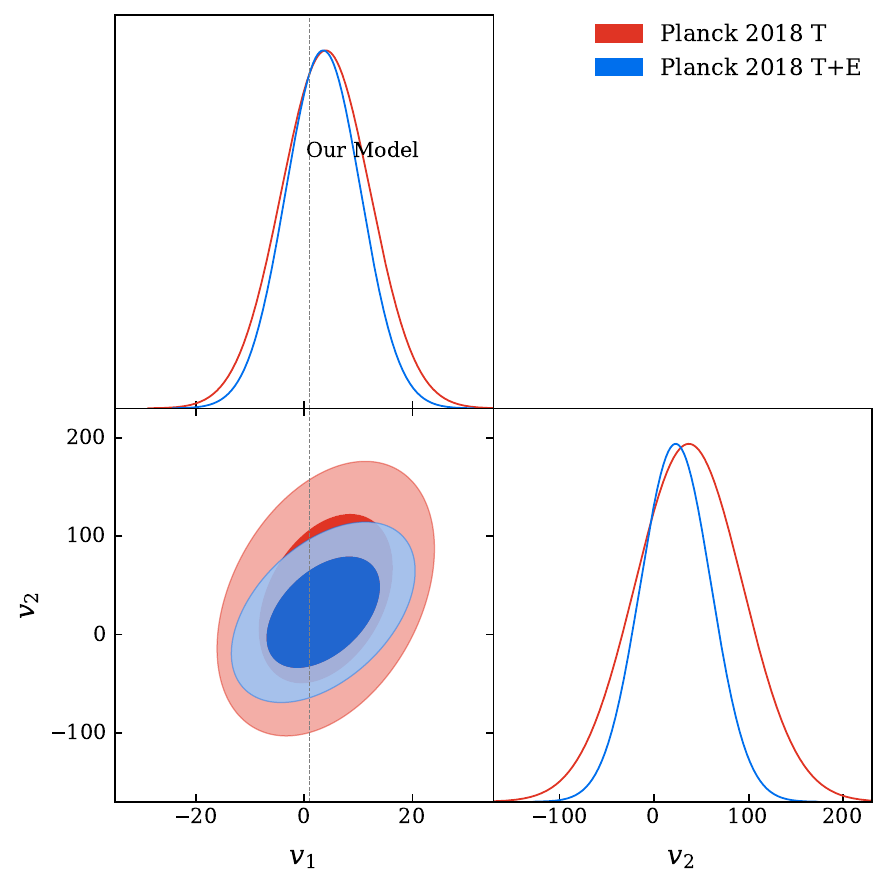}
	\caption{CMB bispectrum constraints on the two effective parameters $v_1$ and $v_2$ as with Figure \ref{fig:triangle_plot}, but with an alternative set of model parameters fixed at the power spectrum level. We find that the model prediction $v_1=1$ is now consistent with the marginalized constraints.}
	\label{fig:triangle_plot_newPS}
\end{figure*}

\begin{figure*}[htbp!]
	\centering	\includegraphics[width=0.65\textwidth]{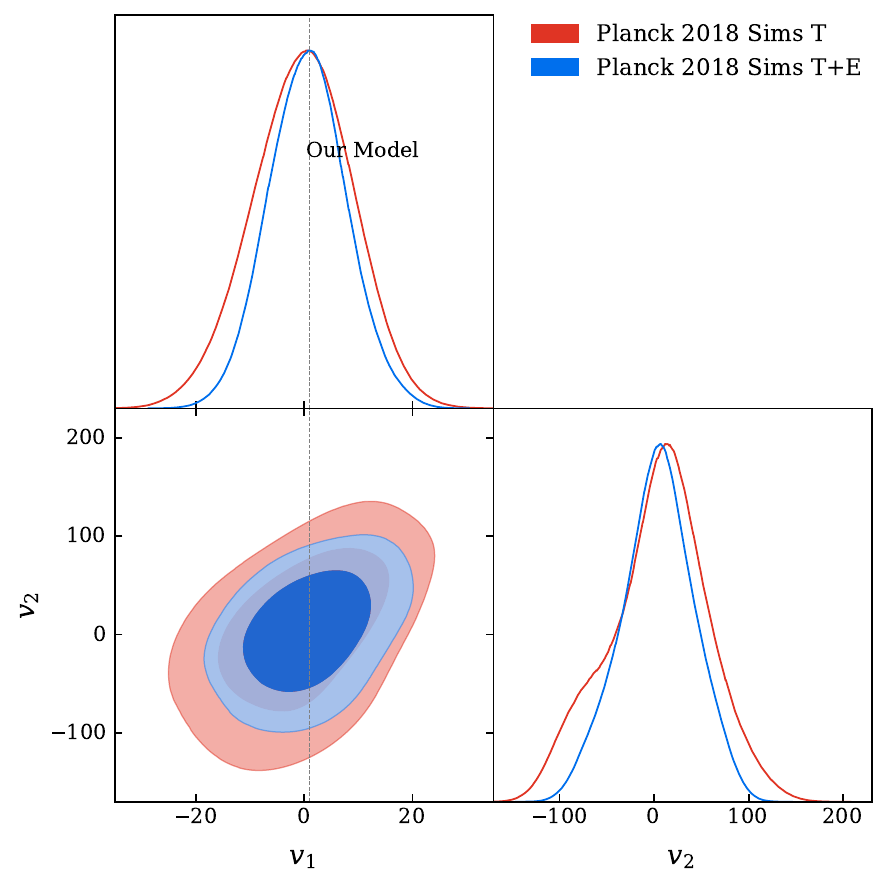}
	\caption{The distribution of the best-fit values of two effective parameters $v_1$ and $v_2$ in the DHOST inflation model for 160 \textit{Planck} end-to-end simulations as in Figure \ref{fig:triangle_plot_samples_newPS}, but with an alternative set of model parameters. We find the model prediction $v_1=1$ is consistent with Gaussian initial conditions and indistinguishable from vanishing PNG with current observations.}
	\label{fig:triangle_plot_samples_newPS}
\end{figure*}

The primordial bispectrum for this alternative set of parameters has a much smaller baseline contribution. Otherwise, the shape of each individual contribution to the bispectrum remains nearly identical as before, so that Figure \ref{fig:bispectrum_shape_plot} is still accurate for this model up to some constant factors. We again parameterize the baseline contribution as $v_1=1$ and plot the CMB constraints as Figures \ref{fig:triangle_plot_newPS} and \ref{fig:triangle_plot_samples_newPS}, which correspond to Figures \ref{fig:triangle_plot} and \ref{fig:triangle_plot_samples}.

Since the shapes of the bispectrum contributions are nearly identical to the previous model, the overall shape of the constraint contours remains the same. However, the amplitude of the baseline bispectrum is now roughly 47 times smaller than before, so the bounds on $v_1$ are now multiplied by 47 and therefore include $v_1=1$ comfortably. Furthermore, the model prediction $v_1=1$ is now consistent with the distribution of $v_1$ and $v_2$ estimated from Gaussian simulations, as seen in Figure \ref{fig:triangle_plot_samples_newPS}; the model is compatible with the current non-detection of PNG.

This shows that DHOST theories can be used to build solid inflationary models, in agreement with both power spectrum and bispectrum constraints. However, as we have shown in this work, the agreement with the non-Gaussian constraints is not obvious without a full analysis of the bispectrum. In both models considered in this work, the bispectrum comprises two distinct shapes from the `base' contribution and the rest that depends linearly on $\bs{\theta}$. The power spectrum parameters affect the amplitude of the base contribution, which can sometimes be too large to be compatible with the observations. Through a full-shape analysis, we confirmed that the second model has $\bs{\theta}$ values that are fully consistent with the \textit{Planck} bispectrum constraints, while the first model does not.

\section{Conclusion}
\label{sec:conclusions}

In this work, we thoroughly investigated the CMB bispectrum constraints on DHOST inflation \cite{Brax2021ps} based on the analytic formulae of its primordial bispectrum obtained in \cite{Brax2021bis}. The expression was converted to \texttt{C} and then wrapped into \texttt{Python} to be compatible with the publicly available code CMB-BEST \cite{Sohn2023cmbbest,Sohn2023cmbbestcode}, which provides \textit{Planck} 2018 constraints on any bispectrum shape.

After enforcing the primordial power spectrum to be consistent with \textit{Planck}, the model contains five explicit free parameters that contribute linearly to the bispectrum. We tried two distinct parameter choices fixed at the power spectrum level and reproduced the fudge factor analyses from previous studies. Through our full-shape analyses of these contributions, we found that there is a fixed bispectrum contribution and one extra degree of freedom in the model, which influence the squeezed and equilateral limits, respectively. Marginalized constraints have been obtained from \textit{Planck} 2018 temperature and polarisation data using the public code CMB-BEST.

The constraints show that one of the two models is disfavored by the data at a $6\sigma$ level. This is mainly due to the large fixed (`base') contribution to the bispectrum in the squeezed limit, which cannot be cancelled out by the model parameters without introducing an even larger bispectrum in other limits. We conclude that this DHOST inflation model, despite being able to explain the near-scale invariant power spectrum and having five free parameters, is not compatible with the \textit{Planck} CMB data.

We emphasize that the conventional `fudge factor' approximation can be misleading and needs to be used with caution. Our example shows that the fudge factors can be set to arbitrarily small amounts by tuning the free parameters, even though the model is ruled out as a whole. A full-shape bispectrum analysis shown in this work is essential for finding accurate constraints in such cases.

While our CMB bispectrum analysis placed new constraints and ruled out a model from previous works, DHOST inflation is still viable. The second model, with alternative parameter choices at the power spectrum level, yields a much smaller bispectrum consistent with the observational limits. We leave the comprehensive study of all DHOST inflation parameters at both the power spectrum and bispectrum level as future work.

In this work, we focused on the DHOST inflation model with the axionic potential studied in \cite{Brax2021ps,Brax2021bis}. Although this should be a fair representative for perturbations to the scale-invariant case, other potentials could generate different types of non-Gaussianities. A full exploration of this possibility is left for future work.

%\newpage

\section*{Acknowledgements}

WS would like to thank Rodrigo Calderon and Arman Shafieloo for useful discussions. This work was supported by a United Kingdom Research and Innovation (UKRI) Future Leaders Fellowship [Grant No.~MR/V021974/2] (AL).

For the purpose of open access, the authors have applied a Creative Commons Attribution (CC BY) licence to any Author Accepted Manuscript version arising.

\section*{Data Access Statement}
The code employed in this work is detailed in Ref. \cite{Sohn2023cmbbest} and available at \cite{Sohn2023cmbbestcode}. The code requires a set of precomputed data for some basis functions. Provided with the code is an HDF5 data file that contains the precomputed results for the \textit{Planck} 2018 CMB temperature and polarization dataset.

\appendix
\section{Consistency with theoretical constraints}

In this section, we derive some theoretical bounds on the DHOST inflation parameters and check their consistency with the CMB bispectrum constraints.

Returning to the dimensionful couplings and using Eqs. (\ref{coordinets-ch}) and (\ref{dimensionless-couplings}), we can express the terms from the action as an effective field theory expansion as
\begin{align}
F_0(X)=\Lambda^4 \sum_{n \ge 1} f_{0,n} \left(\frac{X}{M^2 \Lambda^2} \right)^n \\
F_1(X) \Box \phi=\Lambda^4 \sum_{n \ge 1} f_{1,n} \left(\frac{X}{M^2 \Lambda^2} \right)^n \Box_x \varphi 
\end{align}
where $f_{i,n} \equiv \frac{d^n f_i}{dx^n}$ and $\Box_x$ denotes the box operator with respect to $x$. We perform a rescaling of the variables, assuming that one of the monomials defined by $f_{0,n}$ is large:
\begin{equation}
M^2 \Lambda^2 \to \frac{M^2 \Lambda^2}{f_{0,n}^{1/n}} \, .    
\end{equation}
Under this rescaling, from Eq. (\ref{eq:muc}),
\begin{equation}
\mu_c \to \frac{\mu_c}{f_{0,n}^{1/(4n)}}
\end{equation}
and then 
\begin{equation}
m_{\rm phys}^2=-\frac{\mu^4}{f^2}=-|m^2| \frac{\Lambda^4}{M^2}
\end{equation}
Hence, the condition $m_{\rm phys} \ll \mu_c$ yields
\begin{equation}
M \gg  |m|^{2/3} m_{\rm Pl} f_{0,n}^{1/(6n)} \, ,
\end{equation}
which can be combined with the constraint on the absence of large excursions $M \ll 10^{-6} m_{\rm Pl}$ and gives
\begin{equation}
f_{0,n} \ll \left(\frac{10^{-36}}{|m|^4}\right)^n   \, .
\end{equation}
In our model, $|m^2| \sim 10^{-23}$, hence the constraint becomes $f_{0,n} \ll 10^{10n}$, which is consistent with our constraints.

\bibliography{references}% Produces the bibliography via BibTeX.

%apsrev4-2.bst 2019-01-14 (MD) hand-edited version of apsrev4-1.bst
%Control: key (0)
%Control: author (8) initials jnrlst
%Control: editor formatted (1) identically to author
%Control: production of article title (0) allowed
%Control: page (0) single
%Control: year (1) truncated
%Control: production of eprint (0) enabled
\begin{thebibliography}{48}%
\makeatletter
\providecommand \@ifxundefined [1]{%
 \@ifx{#1\undefined}
}%
\providecommand \@ifnum [1]{%
 \ifnum #1\expandafter \@firstoftwo
 \else \expandafter \@secondoftwo
 \fi
}%
\providecommand \@ifx [1]{%
 \ifx #1\expandafter \@firstoftwo
 \else \expandafter \@secondoftwo
 \fi
}%
\providecommand \natexlab [1]{#1}%
\providecommand \enquote  [1]{``#1''}%
\providecommand \bibnamefont  [1]{#1}%
\providecommand \bibfnamefont [1]{#1}%
\providecommand \citenamefont [1]{#1}%
\providecommand \href@noop [0]{\@secondoftwo}%
\providecommand \href [0]{\begingroup \@sanitize@url \@href}%
\providecommand \@href[1]{\@@startlink{#1}\@@href}%
\providecommand \@@href[1]{\endgroup#1\@@endlink}%
\providecommand \@sanitize@url [0]{\catcode `\\12\catcode `\$12\catcode
  `\&12\catcode `\#12\catcode `\^12\catcode `\_12\catcode `\%12\relax}%
\providecommand \@@startlink[1]{}%
\providecommand \@@endlink[0]{}%
\providecommand \url  [0]{\begingroup\@sanitize@url \@url }%
\providecommand \@url [1]{\endgroup\@href {#1}{\urlprefix }}%
\providecommand \urlprefix  [0]{URL }%
\providecommand \Eprint [0]{\href }%
\providecommand \doibase [0]{https://doi.org/}%
\providecommand \selectlanguage [0]{\@gobble}%
\providecommand \bibinfo  [0]{\@secondoftwo}%
\providecommand \bibfield  [0]{\@secondoftwo}%
\providecommand \translation [1]{[#1]}%
\providecommand \BibitemOpen [0]{}%
\providecommand \bibitemStop [0]{}%
\providecommand \bibitemNoStop [0]{.\EOS\space}%
\providecommand \EOS [0]{\spacefactor3000\relax}%
\providecommand \BibitemShut  [1]{\csname bibitem#1\endcsname}%
\let\auto@bib@innerbib\@empty
%</preamble>
\bibitem [{\citenamefont {{Planck
  Collaboration}}(2020{\natexlab{a}})}]{PlanckCollaboration2018Parameters}%
  \BibitemOpen
  \bibfield  {author} {\bibinfo {author} {\bibnamefont {{Planck
  Collaboration}}},\ }\bibfield  {title} {\bibinfo {title} {{Planck 2018
  results. VI. Cosmological parameters}},\ }\href
  {https://doi.org/10.1051/0004-6361/201833910} {\bibfield  {journal} {\bibinfo
   {journal} {Astronomy {\&} Astrophysics}\ }\textbf {\bibinfo {volume}
  {641}},\ \bibinfo {pages} {A6} (\bibinfo {year}
  {2020}{\natexlab{a}})}\BibitemShut {NoStop}%
\bibitem [{\citenamefont {Guth}(1981)}]{Guth1981inflation}%
  \BibitemOpen
  \bibfield  {author} {\bibinfo {author} {\bibfnamefont {A.~H.}\ \bibnamefont
  {Guth}},\ }\bibfield  {title} {\bibinfo {title} {Inflationary universe: A
  possible solution to the horizon and flatness problems},\ }\href@noop {}
  {\bibfield  {journal} {\bibinfo  {journal} {Physical Review D}\ }\textbf
  {\bibinfo {volume} {23}},\ \bibinfo {pages} {347} (\bibinfo {year}
  {1981})}\BibitemShut {NoStop}%
\bibitem [{\citenamefont {Mukhanov}\ and\ \citenamefont
  {Chibisov}(1981)}]{Mukhanov:1981xt}%
  \BibitemOpen
  \bibfield  {author} {\bibinfo {author} {\bibfnamefont {V.~F.}\ \bibnamefont
  {Mukhanov}}\ and\ \bibinfo {author} {\bibfnamefont {G.~V.}\ \bibnamefont
  {Chibisov}},\ }\bibfield  {title} {\bibinfo {title} {{Quantum Fluctuations
  and a Nonsingular Universe}},\ }\href@noop {} {\bibfield  {journal} {\bibinfo
   {journal} {JETP Lett.}\ }\textbf {\bibinfo {volume} {33}},\ \bibinfo {pages}
  {532} (\bibinfo {year} {1981})}\BibitemShut {NoStop}%
\bibitem [{\citenamefont {Hawking}(1982)}]{Hawking:1982cz}%
  \BibitemOpen
  \bibfield  {author} {\bibinfo {author} {\bibfnamefont {S.~W.}\ \bibnamefont
  {Hawking}},\ }\bibfield  {title} {\bibinfo {title} {{The Development of
  Irregularities in a Single Bubble Inflationary Universe}},\ }\href
  {https://doi.org/10.1016/0370-2693(82)90373-2} {\bibfield  {journal}
  {\bibinfo  {journal} {Physics Letters B}\ }\textbf {\bibinfo {volume}
  {115}},\ \bibinfo {pages} {295} (\bibinfo {year} {1982})}\BibitemShut
  {NoStop}%
\bibitem [{\citenamefont {Kachru}\ \emph {et~al.}(2003)\citenamefont {Kachru},
  \citenamefont {Kallosh}, \citenamefont {Linde}, \citenamefont {Maldacena},
  \citenamefont {McAllister},\ and\ \citenamefont {Trivedi}}]{Kachru:2003sx}%
  \BibitemOpen
  \bibfield  {author} {\bibinfo {author} {\bibfnamefont {S.}~\bibnamefont
  {Kachru}}, \bibinfo {author} {\bibfnamefont {R.}~\bibnamefont {Kallosh}},
  \bibinfo {author} {\bibfnamefont {A.}~\bibnamefont {Linde}}, \bibinfo
  {author} {\bibfnamefont {J.}~\bibnamefont {Maldacena}}, \bibinfo {author}
  {\bibfnamefont {L.}~\bibnamefont {McAllister}},\ and\ \bibinfo {author}
  {\bibfnamefont {S.~P.}\ \bibnamefont {Trivedi}},\ }\bibfield  {title}
  {\bibinfo {title} {Towards inflation in string theory},\ }\href
  {https://doi.org/10.1088/1475-7516/2003/10/013} {\bibfield  {journal}
  {\bibinfo  {journal} {Journal of Cosmology and Astroparticle Physics}\
  }\textbf {\bibinfo {volume} {2003}}\bibinfo  {number} { (10)},\ \bibinfo
  {pages} {013}}\BibitemShut {NoStop}%
\bibitem [{\citenamefont {Baumann}\ and\ \citenamefont
  {McAllister}(2015)}]{baumann_mcallister_2015}%
  \BibitemOpen
\bibfield  {number} {  }\bibfield  {author} {\bibinfo {author} {\bibfnamefont
  {D.}~\bibnamefont {Baumann}}\ and\ \bibinfo {author} {\bibfnamefont
  {L.}~\bibnamefont {McAllister}},\ }\href
  {https://doi.org/10.1017/CBO9781316105733} {\emph {\bibinfo {title}
  {Inflation and String Theory}}},\ Cambridge Monographs on Mathematical
  Physics\ (\bibinfo  {publisher} {Cambridge University Press},\ \bibinfo
  {year} {2015})\BibitemShut {NoStop}%
\bibitem [{\citenamefont {Martin}\ \emph {et~al.}(2014)\citenamefont {Martin},
  \citenamefont {Ringeval},\ and\ \citenamefont {Vennin}}]{Martin:2013tda}%
  \BibitemOpen
  \bibfield  {author} {\bibinfo {author} {\bibfnamefont {J.}~\bibnamefont
  {Martin}}, \bibinfo {author} {\bibfnamefont {C.}~\bibnamefont {Ringeval}},\
  and\ \bibinfo {author} {\bibfnamefont {V.}~\bibnamefont {Vennin}},\
  }\bibfield  {title} {\bibinfo {title} {{Encyclop\ae{}dia Inflationaris}},\
  }\href {https://doi.org/10.1016/j.dark.2014.01.003} {\bibfield  {journal}
  {\bibinfo  {journal} {Phys. Dark Univ.}\ }\textbf {\bibinfo {volume} {5-6}},\
  \bibinfo {pages} {75} (\bibinfo {year} {2014})},\ \Eprint
  {https://arxiv.org/abs/1303.3787} {arXiv:1303.3787 [astro-ph.CO]}
  \BibitemShut {NoStop}%
\bibitem [{\citenamefont {{Planck
  Collaboration}}(2020{\natexlab{b}})}]{PlanckCollaboration2018inflation}%
  \BibitemOpen
  \bibfield  {author} {\bibinfo {author} {\bibnamefont {{Planck
  Collaboration}}},\ }\bibfield  {title} {\bibinfo {title} {{Planck 2018
  results. X. Constraints on inflation}},\ }\href@noop {} {\bibfield  {journal}
  {\bibinfo  {journal} {Astronomy {\&} Astrophysics}\ }\textbf {\bibinfo
  {volume} {641}},\ \bibinfo {pages} {A10} (\bibinfo {year}
  {2020}{\natexlab{b}})}\BibitemShut {NoStop}%
\bibitem [{\citenamefont {Langlois}\ and\ \citenamefont
  {Noui}(2016)}]{Langlois:2015cwa}%
  \BibitemOpen
  \bibfield  {author} {\bibinfo {author} {\bibfnamefont {D.}~\bibnamefont
  {Langlois}}\ and\ \bibinfo {author} {\bibfnamefont {K.}~\bibnamefont
  {Noui}},\ }\bibfield  {title} {\bibinfo {title} {Degenerate higher derivative
  theories beyond horndeski: evading the ostrogradski instability},\ }\href
  {https://doi.org/10.1088/1475-7516/2016/02/034} {\bibfield  {journal}
  {\bibinfo  {journal} {Journal of Cosmology and Astroparticle Physics}\
  }\textbf {\bibinfo {volume} {2016}}\bibinfo  {number} { (02)},\ \bibinfo
  {pages} {034}}\BibitemShut {NoStop}%
\bibitem [{\citenamefont {Ben~Achour}\ \emph {et~al.}(2016)\citenamefont
  {Ben~Achour}, \citenamefont {Langlois},\ and\ \citenamefont
  {Noui}}]{Achour:2016rkg}%
  \BibitemOpen
\bibfield  {number} {  }\bibfield  {author} {\bibinfo {author} {\bibfnamefont
  {J.}~\bibnamefont {Ben~Achour}}, \bibinfo {author} {\bibfnamefont
  {D.}~\bibnamefont {Langlois}},\ and\ \bibinfo {author} {\bibfnamefont
  {K.}~\bibnamefont {Noui}},\ }\bibfield  {title} {\bibinfo {title}
  {{Degenerate higher order scalar-tensor theories beyond Horndeski and
  disformal transformations}},\ }\href
  {https://doi.org/10.1103/PhysRevD.93.124005} {\bibfield  {journal} {\bibinfo
  {journal} {Physical Review D}\ }\textbf {\bibinfo {volume} {93}},\ \bibinfo
  {pages} {124005} (\bibinfo {year} {2016})},\ \Eprint
  {https://arxiv.org/abs/1602.08398} {arXiv:1602.08398 [gr-qc]} \BibitemShut
  {NoStop}%
\bibitem [{\citenamefont {Achour}\ \emph {et~al.}(2016)\citenamefont {Achour},
  \citenamefont {Crisostomi}, \citenamefont {Koyama}, \citenamefont {Langlois},
  \citenamefont {Noui},\ and\ \citenamefont {Tasinato}}]{BenAchour:2016fzp}%
  \BibitemOpen
  \bibfield  {author} {\bibinfo {author} {\bibfnamefont {J.~B.}\ \bibnamefont
  {Achour}}, \bibinfo {author} {\bibfnamefont {M.}~\bibnamefont {Crisostomi}},
  \bibinfo {author} {\bibfnamefont {K.}~\bibnamefont {Koyama}}, \bibinfo
  {author} {\bibfnamefont {D.}~\bibnamefont {Langlois}}, \bibinfo {author}
  {\bibfnamefont {K.}~\bibnamefont {Noui}},\ and\ \bibinfo {author}
  {\bibfnamefont {G.}~\bibnamefont {Tasinato}},\ }\bibfield  {title} {\bibinfo
  {title} {{Degenerate higher order scalar-tensor theories beyond Horndeski up
  to cubic order}},\ }\href {https://doi.org/10.1007/JHEP12(2016)100}
  {\bibfield  {journal} {\bibinfo  {journal} {Journal of High Energy Physics}\
  }\textbf {\bibinfo {volume} {2016}},\ \bibinfo {pages} {100} (\bibinfo {year}
  {2016})}\BibitemShut {NoStop}%
\bibitem [{\citenamefont {Crisostomi}\ \emph {et~al.}(2016)\citenamefont
  {Crisostomi}, \citenamefont {Koyama},\ and\ \citenamefont
  {Tasinato}}]{Crisostomi:2016czh}%
  \BibitemOpen
  \bibfield  {author} {\bibinfo {author} {\bibfnamefont {M.}~\bibnamefont
  {Crisostomi}}, \bibinfo {author} {\bibfnamefont {K.}~\bibnamefont {Koyama}},\
  and\ \bibinfo {author} {\bibfnamefont {G.}~\bibnamefont {Tasinato}},\
  }\bibfield  {title} {\bibinfo {title} {Extended scalar-tensor theories of
  gravity},\ }\href {https://doi.org/10.1088/1475-7516/2016/04/044} {\bibfield
  {journal} {\bibinfo  {journal} {Journal of Cosmology and Astroparticle
  Physics}\ }\textbf {\bibinfo {volume} {2016}}\bibinfo  {number} { (04)},\
  \bibinfo {pages} {044}}\BibitemShut {NoStop}%
\bibitem [{\citenamefont {Crisostomi}\ \emph
  {et~al.}(2019{\natexlab{a}})\citenamefont {Crisostomi}, \citenamefont
  {Koyama}, \citenamefont {Langlois}, \citenamefont {Noui},\ and\ \citenamefont
  {Steer}}]{Crisostomi:2018bsp}%
  \BibitemOpen
\bibfield  {number} {  }\bibfield  {author} {\bibinfo {author} {\bibfnamefont
  {M.}~\bibnamefont {Crisostomi}}, \bibinfo {author} {\bibfnamefont
  {K.}~\bibnamefont {Koyama}}, \bibinfo {author} {\bibfnamefont
  {D.}~\bibnamefont {Langlois}}, \bibinfo {author} {\bibfnamefont
  {K.}~\bibnamefont {Noui}},\ and\ \bibinfo {author} {\bibfnamefont
  {D.}~\bibnamefont {Steer}},\ }\bibfield  {title} {\bibinfo {title}
  {{Cosmological evolution in DHOST theories}},\ }\href
  {https://doi.org/10.1088/1475-7516/2019/01/030} {\bibfield  {journal}
  {\bibinfo  {journal} {Journal of Cosmology and Astroparticle Physics}\
  }\textbf {\bibinfo {volume} {2019}}\bibinfo  {number} { (01)},\ \bibinfo
  {pages} {030}}\BibitemShut {NoStop}%
\bibitem [{\citenamefont {Bombacigno}\ \emph {et~al.}(2021)\citenamefont
  {Bombacigno}, \citenamefont {Boudet}, \citenamefont {Olmo},\ and\
  \citenamefont {Montani}}]{Bombacigno:2021bpk}%
  \BibitemOpen
\bibfield  {number} {  }\bibfield  {author} {\bibinfo {author} {\bibfnamefont
  {F.}~\bibnamefont {Bombacigno}}, \bibinfo {author} {\bibfnamefont
  {S.}~\bibnamefont {Boudet}}, \bibinfo {author} {\bibfnamefont {G.~J.}\
  \bibnamefont {Olmo}},\ and\ \bibinfo {author} {\bibfnamefont
  {G.}~\bibnamefont {Montani}},\ }\bibfield  {title} {\bibinfo {title} {{Big
  bounce and future time singularity resolution in Bianchi I cosmologies: The
  projective invariant Nieh-Yan case}},\ }\href
  {https://doi.org/10.1103/PhysRevD.103.124031} {\bibfield  {journal} {\bibinfo
   {journal} {Physical Review D}\ }\textbf {\bibinfo {volume} {103}},\ \bibinfo
  {pages} {124031} (\bibinfo {year} {2021})},\ \Eprint
  {https://arxiv.org/abs/2105.06870} {arXiv:2105.06870 [gr-qc]} \BibitemShut
  {NoStop}%
\bibitem [{\citenamefont {Ostrogradsky}(1850)}]{Ostrogradsky:1850fid}%
  \BibitemOpen
  \bibfield  {author} {\bibinfo {author} {\bibfnamefont {M.}~\bibnamefont
  {Ostrogradsky}},\ }\bibfield  {title} {\bibinfo {title} {{M\'emoires sur les
  \'equations diff\'erentielles, relatives au probl\`eme des
  isop\'erim\`etres}},\ }\href@noop {} {\bibfield  {journal} {\bibinfo
  {journal} {Mem. Acad. St. Petersbourg}\ }\textbf {\bibinfo {volume} {6}},\
  \bibinfo {pages} {385} (\bibinfo {year} {1850})}\BibitemShut {NoStop}%
\bibitem [{\citenamefont {{Horndeski}}(1974)}]{Horndeski1974}%
  \BibitemOpen
  \bibfield  {author} {\bibinfo {author} {\bibfnamefont {G.~W.}\ \bibnamefont
  {{Horndeski}}},\ }\bibfield  {title} {\bibinfo {title} {{Second-Order
  Scalar-Tensor Field Equations in a Four-Dimensional Space}},\ }\href
  {https://doi.org/10.1007/BF01807638} {\bibfield  {journal} {\bibinfo
  {journal} {International Journal of Theoretical Physics}\ }\textbf {\bibinfo
  {volume} {10}},\ \bibinfo {pages} {363} (\bibinfo {year} {1974})}\BibitemShut
  {NoStop}%
\bibitem [{\citenamefont {Zumalac\'arregui}\ and\ \citenamefont
  {Garc\'\i{}a-Bellido}(2014)}]{Zumalacarregui:2013pma}%
  \BibitemOpen
  \bibfield  {author} {\bibinfo {author} {\bibfnamefont {M.}~\bibnamefont
  {Zumalac\'arregui}}\ and\ \bibinfo {author} {\bibfnamefont {J.}~\bibnamefont
  {Garc\'\i{}a-Bellido}},\ }\bibfield  {title} {\bibinfo {title} {{Transforming
  gravity: from derivative couplings to matter to second-order scalar-tensor
  theories beyond the Horndeski Lagrangian}},\ }\href
  {https://doi.org/10.1103/PhysRevD.89.064046} {\bibfield  {journal} {\bibinfo
  {journal} {Physical Review D}\ }\textbf {\bibinfo {volume} {89}},\ \bibinfo
  {pages} {064046} (\bibinfo {year} {2014})},\ \Eprint
  {https://arxiv.org/abs/1308.4685} {arXiv:1308.4685 [gr-qc]} \BibitemShut
  {NoStop}%
\bibitem [{\citenamefont {Gleyzes}\ \emph
  {et~al.}(2015{\natexlab{a}})\citenamefont {Gleyzes}, \citenamefont
  {Langlois}, \citenamefont {Piazza},\ and\ \citenamefont
  {Vernizzi}}]{Gleyzes:2014dya}%
  \BibitemOpen
  \bibfield  {author} {\bibinfo {author} {\bibfnamefont {J.}~\bibnamefont
  {Gleyzes}}, \bibinfo {author} {\bibfnamefont {D.}~\bibnamefont {Langlois}},
  \bibinfo {author} {\bibfnamefont {F.}~\bibnamefont {Piazza}},\ and\ \bibinfo
  {author} {\bibfnamefont {F.}~\bibnamefont {Vernizzi}},\ }\bibfield  {title}
  {\bibinfo {title} {{Healthy theories beyond Horndeski}},\ }\href
  {https://doi.org/10.1103/PhysRevLett.114.211101} {\bibfield  {journal}
  {\bibinfo  {journal} {Phys. Rev. Lett.}\ }\textbf {\bibinfo {volume} {114}},\
  \bibinfo {pages} {211101} (\bibinfo {year} {2015}{\natexlab{a}})},\ \Eprint
  {https://arxiv.org/abs/1404.6495} {arXiv:1404.6495 [hep-th]} \BibitemShut
  {NoStop}%
\bibitem [{\citenamefont {Gleyzes}\ \emph
  {et~al.}(2015{\natexlab{b}})\citenamefont {Gleyzes}, \citenamefont
  {Langlois}, \citenamefont {Piazza},\ and\ \citenamefont
  {Vernizzi}}]{Gleyzes:2014qga}%
  \BibitemOpen
  \bibfield  {author} {\bibinfo {author} {\bibfnamefont {J.}~\bibnamefont
  {Gleyzes}}, \bibinfo {author} {\bibfnamefont {D.}~\bibnamefont {Langlois}},
  \bibinfo {author} {\bibfnamefont {F.}~\bibnamefont {Piazza}},\ and\ \bibinfo
  {author} {\bibfnamefont {F.}~\bibnamefont {Vernizzi}},\ }\bibfield  {title}
  {\bibinfo {title} {Exploring gravitational theories beyond horndeski},\
  }\href {https://dx.doi.org/10.1088/1475-7516/2015/02/018} {\bibfield
  {journal} {\bibinfo  {journal} {Journal of Cosmology and Astroparticle
  Physics}\ }\textbf {\bibinfo {volume} {2015}}\bibinfo  {number} { (02)},\
  \bibinfo {pages} {018}}\BibitemShut {NoStop}%
\bibitem [{\citenamefont {Brax}\ and\ \citenamefont
  {Lazanu}(2021)}]{Brax2021ps}%
  \BibitemOpen
\bibfield  {number} {  }\bibfield  {author} {\bibinfo {author} {\bibfnamefont
  {P.}~\bibnamefont {Brax}}\ and\ \bibinfo {author} {\bibfnamefont
  {A.}~\bibnamefont {Lazanu}},\ }\bibfield  {title} {\bibinfo {title}
  {{Scale-dependence in DHOST inflation}},\ }\href
  {https://doi.org/10.1088/1475-7516/2021/08/061} {\bibfield  {journal}
  {\bibinfo  {journal} {Journal of Cosmology and Astroparticle Physics}\
  }\textbf {\bibinfo {volume} {2021}}\bibinfo  {number} { (08)},\ \bibinfo
  {pages} {061}}\BibitemShut {NoStop}%
\bibitem [{\citenamefont {Brax}\ and\ \citenamefont
  {Lazanu}(2022)}]{Brax2021bis}%
  \BibitemOpen
\bibfield  {number} {  }\bibfield  {author} {\bibinfo {author} {\bibfnamefont
  {P.}~\bibnamefont {Brax}}\ and\ \bibinfo {author} {\bibfnamefont
  {A.}~\bibnamefont {Lazanu}},\ }\bibfield  {title} {\bibinfo {title}
  {{Non-Gaussianity in DHOST inflation}},\ }\href
  {https://doi.org/10.1088/1475-7516/2022/01/026} {\bibfield  {journal}
  {\bibinfo  {journal} {Journal of Cosmology and Astroparticle Physics}\
  }\textbf {\bibinfo {volume} {01}}\bibfield  {number} {\bibinfo  {number} {
  (01)},\ \bibinfo {pages} {026}},\ }\Eprint {https://arxiv.org/abs/2110.05913}
  {arXiv:2110.05913 [astro-ph.CO]} \BibitemShut {NoStop}%
\bibitem [{\citenamefont {{Planck
  Collaboration}}(2020{\natexlab{c}})}]{PlanckCollaboration2018}%
  \BibitemOpen
  \bibfield  {author} {\bibinfo {author} {\bibnamefont {{Planck
  Collaboration}}},\ }\bibfield  {title} {\bibinfo {title} {{Planck 2018
  results. IX. Constraints on primordial non-Gaussianity}},\ }\href
  {https://doi.org/10.1051/0004-6361/201935891} {\bibfield  {journal} {\bibinfo
   {journal} {Astronomy {\&} Astrophysics}\ }\textbf {\bibinfo {volume}
  {641}},\ \bibinfo {pages} {A9} (\bibinfo {year}
  {2020}{\natexlab{c}})}\BibitemShut {NoStop}%
\bibitem [{\citenamefont {Sohn}\ \emph {et~al.}(2023)\citenamefont {Sohn},
  \citenamefont {Fergusson},\ and\ \citenamefont {Shellard}}]{Sohn2023cmbbest}%
  \BibitemOpen
  \bibfield  {author} {\bibinfo {author} {\bibfnamefont {W.}~\bibnamefont
  {Sohn}}, \bibinfo {author} {\bibfnamefont {J.~R.}\ \bibnamefont
  {Fergusson}},\ and\ \bibinfo {author} {\bibfnamefont {E.~P.~S.}\ \bibnamefont
  {Shellard}},\ }\bibfield  {title} {\bibinfo {title} {{High-resolution CMB
  bispectrum estimator with flexible modal bases}},\ }\href
  {https://doi.org/10.1103/PhysRevD.108.063504} {\bibfield  {journal} {\bibinfo
   {journal} {Physical Review D}\ }\textbf {\bibinfo {volume} {108}},\ \bibinfo
  {pages} {063504} (\bibinfo {year} {2023})}\BibitemShut {NoStop}%
\bibitem [{\citenamefont {Sohn}(2023)}]{Sohn2023cmbbestcode}%
  \BibitemOpen
  \bibfield  {author} {\bibinfo {author} {\bibfnamefont {W.}~\bibnamefont
  {Sohn}},\ }\href {https://github.com/Wuhyun/CMB-BEST} {\bibinfo {title}
  {{CMB-BEST}}},\ \bibinfo {howpublished} {https://github.com/Wuhyun/CMB-BEST}
  (\bibinfo {year} {2023})\BibitemShut {NoStop}%
\bibitem [{\citenamefont {Crisostomi}\ and\ \citenamefont
  {Koyama}(2018)}]{Crisostomi:2017}%
  \BibitemOpen
  \bibfield  {author} {\bibinfo {author} {\bibfnamefont {M.}~\bibnamefont
  {Crisostomi}}\ and\ \bibinfo {author} {\bibfnamefont {K.}~\bibnamefont
  {Koyama}},\ }\bibfield  {title} {\bibinfo {title} {{Self-accelerating
  universe in scalar-tensor theories after GW170817}},\ }\href
  {https://doi.org/10.1103/PhysRevD.97.084004} {\bibfield  {journal} {\bibinfo
  {journal} {Physical Review D}\ }\textbf {\bibinfo {volume} {97}},\ \bibinfo
  {pages} {084004} (\bibinfo {year} {2018})},\ \Eprint
  {https://arxiv.org/abs/1712.06556} {arXiv:1712.06556 [astro-ph.CO]}
  \BibitemShut {NoStop}%
\bibitem [{\citenamefont {Crisostomi}\ \emph
  {et~al.}(2019{\natexlab{b}})\citenamefont {Crisostomi}, \citenamefont
  {Lewandowski},\ and\ \citenamefont {Vernizzi}}]{Crisostomi:2019}%
  \BibitemOpen
  \bibfield  {author} {\bibinfo {author} {\bibfnamefont {M.}~\bibnamefont
  {Crisostomi}}, \bibinfo {author} {\bibfnamefont {M.}~\bibnamefont
  {Lewandowski}},\ and\ \bibinfo {author} {\bibfnamefont {F.}~\bibnamefont
  {Vernizzi}},\ }\bibfield  {title} {\bibinfo {title} {{Vainshtein regime in
  scalar-tensor gravity: Constraints on degenerate higher-order scalar-tensor
  theories}},\ }\href {https://doi.org/10.1103/PhysRevD.100.024025} {\bibfield
  {journal} {\bibinfo  {journal} {Physical Review D}\ }\textbf {\bibinfo
  {volume} {100}},\ \bibinfo {pages} {024025} (\bibinfo {year}
  {2019}{\natexlab{b}})},\ \Eprint {https://arxiv.org/abs/1903.11591}
  {arXiv:1903.11591 [gr-qc]} \BibitemShut {NoStop}%
\bibitem [{\citenamefont {Marsh}(2016)}]{Marsh2015}%
  \BibitemOpen
  \bibfield  {author} {\bibinfo {author} {\bibfnamefont {D.~J.~E.}\
  \bibnamefont {Marsh}},\ }\bibfield  {title} {\bibinfo {title} {{Axion
  Cosmology}},\ }\href {https://doi.org/10.1016/j.physrep.2016.06.005}
  {\bibfield  {journal} {\bibinfo  {journal} {Phys. Rept.}\ }\textbf {\bibinfo
  {volume} {643}},\ \bibinfo {pages} {1} (\bibinfo {year} {2016})},\ \Eprint
  {https://arxiv.org/abs/1510.07633} {arXiv:1510.07633 [astro-ph.CO]}
  \BibitemShut {NoStop}%
\bibitem [{\citenamefont {Vafa}(2005)}]{Vafa:2005ui}%
  \BibitemOpen
  \bibfield  {author} {\bibinfo {author} {\bibfnamefont {C.}~\bibnamefont
  {Vafa}},\ }\bibfield  {title} {\bibinfo {title} {{{The String landscape and
  the swampland}}},\ }\href@noop {} {\bibfield  {journal} {\bibinfo  {journal}
  {arXiv preprint}\ } (\bibinfo {year} {2005})}\BibitemShut {NoStop}%
\bibitem [{\citenamefont {Bedroya}\ and\ \citenamefont
  {Vafa}(2020)}]{Bedroya:2019snp}%
  \BibitemOpen
  \bibfield  {author} {\bibinfo {author} {\bibfnamefont {A.}~\bibnamefont
  {Bedroya}}\ and\ \bibinfo {author} {\bibfnamefont {C.}~\bibnamefont {Vafa}},\
  }\bibfield  {title} {\bibinfo {title} {{Trans-Planckian Censorship and the
  Swampland}},\ }\href {https://doi.org/10.1007/JHEP09(2020)123} {\bibfield
  {journal} {\bibinfo  {journal} {Journal of High Energy Physics}\ }\textbf
  {\bibinfo {volume} {2020}},\ \bibinfo {pages} {123} (\bibinfo {year}
  {2020})}\BibitemShut {NoStop}%
\bibitem [{\citenamefont {Ooguri}\ and\ \citenamefont
  {Vafa}(2007)}]{Ooguri:2006in}%
  \BibitemOpen
  \bibfield  {author} {\bibinfo {author} {\bibfnamefont {H.}~\bibnamefont
  {Ooguri}}\ and\ \bibinfo {author} {\bibfnamefont {C.}~\bibnamefont {Vafa}},\
  }\bibfield  {title} {\bibinfo {title} {{On the Geometry of the String
  Landscape and the Swampland}},\ }\href
  {https://doi.org/10.1016/j.nuclphysb.2006.10.033} {\bibfield  {journal}
  {\bibinfo  {journal} {Nucl. Phys. B}\ }\textbf {\bibinfo {volume} {766}},\
  \bibinfo {pages} {21} (\bibinfo {year} {2007})},\ \Eprint
  {https://arxiv.org/abs/hep-th/0605264} {arXiv:hep-th/0605264} \BibitemShut
  {NoStop}%
\bibitem [{\citenamefont {Gorji}\ \emph {et~al.}(2021)\citenamefont {Gorji},
  \citenamefont {Motohashi},\ and\ \citenamefont {Mukohyama}}]{Gorji:2020bfl}%
  \BibitemOpen
  \bibfield  {author} {\bibinfo {author} {\bibfnamefont {M.~A.}\ \bibnamefont
  {Gorji}}, \bibinfo {author} {\bibfnamefont {H.}~\bibnamefont {Motohashi}},\
  and\ \bibinfo {author} {\bibfnamefont {S.}~\bibnamefont {Mukohyama}},\
  }\bibfield  {title} {\bibinfo {title} {{Stealth dark energy in scordatura
  DHOST theory}},\ }\href {https://doi.org/10.1088/1475-7516/2021/03/081}
  {\bibfield  {journal} {\bibinfo  {journal} {Journal of Cosmology and
  Astroparticle Physics}\ }\textbf {\bibinfo {volume} {2021}}\bibinfo  {number}
  { (03)},\ \bibinfo {pages} {081}}\BibitemShut {NoStop}%
\bibitem [{\citenamefont {Peter}\ and\ \citenamefont
  {Uzan}(2013)}]{Peter:2013avv}%
  \BibitemOpen
\bibfield  {number} {  }\bibfield  {author} {\bibinfo {author} {\bibfnamefont
  {P.}~\bibnamefont {Peter}}\ and\ \bibinfo {author} {\bibfnamefont {J.-P.}\
  \bibnamefont {Uzan}},\ }\href@noop {} {\emph {\bibinfo {title} {{Primordial
  Cosmology}}}},\ Oxford Graduate Texts\ (\bibinfo  {publisher} {Oxford
  University Press},\ \bibinfo {year} {2013})\BibitemShut {NoStop}%
\bibitem [{\citenamefont {Maldacena}(2002)}]{Maldacena2003}%
  \BibitemOpen
  \bibfield  {author} {\bibinfo {author} {\bibfnamefont {J.}~\bibnamefont
  {Maldacena}},\ }\bibfield  {title} {\bibinfo {title} {{Non-Gaussian features
  of primordial fluctuations in single field inflationary models}},\ }\href
  {http://arxiv.org/abs/astro-ph/0210603{\%}0Ahttp://dx.doi.org/10.1088/1126-6708/2003/05/013}
  {\bibfield  {journal} {\bibinfo  {journal} {Journal of High Energy Physics}\
  }\textbf {\bibinfo {volume} {2003}},\ \bibinfo {pages} {13} (\bibinfo {year}
  {2002})},\ \Eprint {https://arxiv.org/abs/0210603} {arXiv:0210603 [astro-ph]}
  \BibitemShut {NoStop}%
\bibitem [{\citenamefont {Chen}(2010)}]{Chen2010review}%
  \BibitemOpen
  \bibfield  {author} {\bibinfo {author} {\bibfnamefont {X.}~\bibnamefont
  {Chen}},\ }\bibfield  {title} {\bibinfo {title} {{Primordial
  Non-Gaussianities from Inflation Models}},\ }\href
  {https://doi.org/10.1155/2010/638979} {\bibfield  {journal} {\bibinfo
  {journal} {Adv. Astron.}\ }\textbf {\bibinfo {volume} {2010}},\ \bibinfo
  {pages} {638979} (\bibinfo {year} {2010})},\ \Eprint
  {https://arxiv.org/abs/1002.1416} {arXiv:1002.1416 [astro-ph.CO]}
  \BibitemShut {NoStop}%
\bibitem [{\citenamefont {Bellini}\ and\ \citenamefont
  {Sawicki}(2014)}]{Bellini:2014fua}%
  \BibitemOpen
  \bibfield  {author} {\bibinfo {author} {\bibfnamefont {E.}~\bibnamefont
  {Bellini}}\ and\ \bibinfo {author} {\bibfnamefont {I.}~\bibnamefont
  {Sawicki}},\ }\bibfield  {title} {\bibinfo {title} {{Maximal freedom at
  minimum cost: linear large-scale structure in general modifications of
  gravity}},\ }\href {https://doi.org/10.1088/1475-7516/2014/07/050} {\bibfield
   {journal} {\bibinfo  {journal} {Journal of Cosmology and Astroparticle
  Physics}\ }\textbf {\bibinfo {volume} {2014}}\bibinfo  {number} { (07)},\
  \bibinfo {pages} {050}}\BibitemShut {NoStop}%
\bibitem [{\citenamefont {Gleyzes}\ \emph
  {et~al.}(2015{\natexlab{c}})\citenamefont {Gleyzes}, \citenamefont
  {Langlois},\ and\ \citenamefont {Vernizzi}}]{Gleyzes:2014rba}%
  \BibitemOpen
\bibfield  {number} {  }\bibfield  {author} {\bibinfo {author} {\bibfnamefont
  {J.}~\bibnamefont {Gleyzes}}, \bibinfo {author} {\bibfnamefont
  {D.}~\bibnamefont {Langlois}},\ and\ \bibinfo {author} {\bibfnamefont
  {F.}~\bibnamefont {Vernizzi}},\ }\bibfield  {title} {\bibinfo {title} {{A
  unifying description of dark energy}},\ }\href
  {https://doi.org/10.1142/S021827181443010X} {\bibfield  {journal} {\bibinfo
  {journal} {Int. J. Mod. Phys. D}\ }\textbf {\bibinfo {volume} {23}},\
  \bibinfo {pages} {1443010} (\bibinfo {year} {2015}{\natexlab{c}})},\ \Eprint
  {https://arxiv.org/abs/1411.3712} {arXiv:1411.3712 [hep-th]} \BibitemShut
  {NoStop}%
\bibitem [{\citenamefont {Langlois}\ \emph {et~al.}(2017)\citenamefont
  {Langlois}, \citenamefont {Mancarella}, \citenamefont {Noui},\ and\
  \citenamefont {Vernizzi}}]{Langlois:2017mxy}%
  \BibitemOpen
  \bibfield  {author} {\bibinfo {author} {\bibfnamefont {D.}~\bibnamefont
  {Langlois}}, \bibinfo {author} {\bibfnamefont {M.}~\bibnamefont
  {Mancarella}}, \bibinfo {author} {\bibfnamefont {K.}~\bibnamefont {Noui}},\
  and\ \bibinfo {author} {\bibfnamefont {F.}~\bibnamefont {Vernizzi}},\
  }\bibfield  {title} {\bibinfo {title} {{Effective description of higher-order
  scalar-tensor theories}},\ }\href
  {https://doi.org/10.1088/1475-7516/2017/05/033} {\bibfield  {journal}
  {\bibinfo  {journal} {Journal of Cosmology and Astroparticle Physics}\
  }\textbf {\bibinfo {volume} {2017}}\bibinfo  {number} { (05)},\ \bibinfo
  {pages} {033}}\BibitemShut {NoStop}%
\bibitem [{\citenamefont {Motohashi}\ and\ \citenamefont
  {Hu}(2017)}]{Motohashi:2017gqb}%
  \BibitemOpen
\bibfield  {number} {  }\bibfield  {author} {\bibinfo {author} {\bibfnamefont
  {H.}~\bibnamefont {Motohashi}}\ and\ \bibinfo {author} {\bibfnamefont
  {W.}~\bibnamefont {Hu}},\ }\bibfield  {title} {\bibinfo {title} {{Generalized
  Slow Roll in the Unified Effective Field Theory of Inflation}},\ }\href
  {https://doi.org/10.1103/PhysRevD.96.023502} {\bibfield  {journal} {\bibinfo
  {journal} {Physical Review D}\ }\textbf {\bibinfo {volume} {96}},\ \bibinfo
  {pages} {023502} (\bibinfo {year} {2017})},\ \Eprint
  {https://arxiv.org/abs/1704.01128} {arXiv:1704.01128 [hep-th]} \BibitemShut
  {NoStop}%
\bibitem [{\citenamefont {Babich}\ \emph {et~al.}(2004)\citenamefont {Babich},
  \citenamefont {Creminelli},\ and\ \citenamefont
  {Zaldarriaga}}]{Babich:2004gb}%
  \BibitemOpen
  \bibfield  {author} {\bibinfo {author} {\bibfnamefont {D.}~\bibnamefont
  {Babich}}, \bibinfo {author} {\bibfnamefont {P.}~\bibnamefont {Creminelli}},\
  and\ \bibinfo {author} {\bibfnamefont {M.}~\bibnamefont {Zaldarriaga}},\
  }\bibfield  {title} {\bibinfo {title} {{The shape of non-Gaussianities}},\
  }\href {https://doi.org/10.1088/1475-7516/2004/08/009} {\bibfield  {journal}
  {\bibinfo  {journal} {Journal of Cosmology and Astroparticle Physics}\
  }\textbf {\bibinfo {volume} {2004}}\bibinfo  {number} { (08)},\ \bibinfo
  {pages} {009}}\BibitemShut {NoStop}%
\bibitem [{\citenamefont {Komatsu}(2010)}]{Komatsu2010}%
  \BibitemOpen
\bibfield  {number} {  }\bibfield  {author} {\bibinfo {author} {\bibfnamefont
  {E.}~\bibnamefont {Komatsu}},\ }\bibfield  {title} {\bibinfo {title}
  {{Hunting for Primordial Non-Gaussianity in the Cosmic Microwave
  Background}},\ }\href
  {http://arxiv.org/abs/1003.6097{\%}0Ahttp://dx.doi.org/10.1088/0264-9381/27/12/124010}
  {\bibfield  {journal} {\bibinfo  {journal} {Classical and Quantum Gravity}\
  }\textbf {\bibinfo {volume} {27}} (\bibinfo {year} {2010})},\ \Eprint
  {https://arxiv.org/abs/1003.6097} {arXiv:1003.6097} \BibitemShut {NoStop}%
\bibitem [{\citenamefont {Liguori}\ \emph {et~al.}(2010)\citenamefont
  {Liguori}, \citenamefont {Sefusatti}, \citenamefont {Fergusson},\ and\
  \citenamefont {Shellard}}]{Liguori2010}%
  \BibitemOpen
  \bibfield  {author} {\bibinfo {author} {\bibfnamefont {M.}~\bibnamefont
  {Liguori}}, \bibinfo {author} {\bibfnamefont {E.}~\bibnamefont {Sefusatti}},
  \bibinfo {author} {\bibfnamefont {J.~R.}\ \bibnamefont {Fergusson}},\ and\
  \bibinfo {author} {\bibfnamefont {E.~P.}\ \bibnamefont {Shellard}},\
  }\bibfield  {title} {\bibinfo {title} {{Primordial non-gaussianity and
  bispectrum measurements in the cosmic microwave background and large-scale
  structure}},\ }\href@noop {} {\bibfield  {journal} {\bibinfo  {journal}
  {Advances in Astronomy}\ }\textbf {\bibinfo {volume} {2010}} (\bibinfo {year}
  {2010})},\ \Eprint {https://arxiv.org/abs/1001.4707} {arXiv:1001.4707}
  \BibitemShut {NoStop}%
\bibitem [{\citenamefont {Fergusson}\ \emph {et~al.}(2012)\citenamefont
  {Fergusson}, \citenamefont {Liguori},\ and\ \citenamefont
  {Shellard}}]{Fergusson2012}%
  \BibitemOpen
  \bibfield  {author} {\bibinfo {author} {\bibfnamefont {J.~R.}\ \bibnamefont
  {Fergusson}}, \bibinfo {author} {\bibfnamefont {M.}~\bibnamefont {Liguori}},\
  and\ \bibinfo {author} {\bibfnamefont {E.~P.~S.}\ \bibnamefont {Shellard}},\
  }\bibfield  {title} {\bibinfo {title} {{The CMB bispectrum}},\ }\href@noop {}
  {\bibfield  {journal} {\bibinfo  {journal} {Journal of Cosmology and
  Astroparticle Physics}\ }\textbf {\bibinfo {volume} {2012}}\bibfield
  {number} {\bibinfo  {number} { (12)}},\ }\Eprint
  {https://arxiv.org/abs/1006.1642} {arXiv:1006.1642} \BibitemShut {NoStop}%
\bibitem [{\citenamefont {{Planck
  Collaboration}}(2016)}]{PlanckCollaboration2015simulations}%
  \BibitemOpen
  \bibfield  {author} {\bibinfo {author} {\bibnamefont {{Planck
  Collaboration}}},\ }\bibfield  {title} {\bibinfo {title} {{Planck 2015
  results. XII. Full Focal Plane Simulations}},\ }\href@noop {} {\bibfield
  {journal} {\bibinfo  {journal} {Astronomy {\&} Astrophysics}\ }\textbf
  {\bibinfo {volume} {594}},\ \bibinfo {pages} {A12} (\bibinfo {year}
  {2016})}\BibitemShut {NoStop}%
\bibitem [{\citenamefont {{Planck
  Collaboration}}(2020{\natexlab{d}})}]{PlanckCollaboration2018hfi}%
  \BibitemOpen
  \bibfield  {author} {\bibinfo {author} {\bibnamefont {{Planck
  Collaboration}}},\ }\bibfield  {title} {\bibinfo {title} {{Planck 2018
  results. III. High Frequency Instrument data processing and frequency
  maps}},\ }\href@noop {} {\bibfield  {journal} {\bibinfo  {journal} {Astronomy
  {\&} Astrophysics}\ }\textbf {\bibinfo {volume} {641}},\ \bibinfo {pages}
  {A3} (\bibinfo {year} {2020}{\natexlab{d}})}\BibitemShut {NoStop}%
\bibitem [{\citenamefont {Garcia-Saenz}\ \emph {et~al.}(2018)\citenamefont
  {Garcia-Saenz}, \citenamefont {Renaux-Petel},\ and\ \citenamefont
  {Ronayne}}]{GarciaSaenz2018fudge}%
  \BibitemOpen
  \bibfield  {author} {\bibinfo {author} {\bibfnamefont {S.}~\bibnamefont
  {Garcia-Saenz}}, \bibinfo {author} {\bibfnamefont {S.}~\bibnamefont
  {Renaux-Petel}},\ and\ \bibinfo {author} {\bibfnamefont {J.}~\bibnamefont
  {Ronayne}},\ }\bibfield  {title} {\bibinfo {title} {{Primordial fluctuations
  and non-Gaussianities in sidetracked inflation}},\ }\href
  {https://doi.org/10.1088/1475-7516/2018/07/057} {\bibfield  {journal}
  {\bibinfo  {journal} {Journal of Cosmology and Astroparticle Physics}\
  }\textbf {\bibinfo {volume} {2018}}\bibinfo  {number} { (07)},\ \bibinfo
  {pages} {057}}\BibitemShut {NoStop}%
\bibitem [{\citenamefont {Koshelev}\ \emph {et~al.}(2023)\citenamefont
  {Koshelev}, \citenamefont {Kumar},\ and\ \citenamefont
  {Starobinsky}}]{Koshelev2023fudge}%
  \BibitemOpen
\bibfield  {number} {  }\bibfield  {author} {\bibinfo {author} {\bibfnamefont
  {A.~S.}\ \bibnamefont {Koshelev}}, \bibinfo {author} {\bibfnamefont {K.~S.}\
  \bibnamefont {Kumar}},\ and\ \bibinfo {author} {\bibfnamefont {A.~A.}\
  \bibnamefont {Starobinsky}},\ }\bibfield  {title} {\bibinfo {title}
  {{Non-Gaussianities in generalized non-local $R^2$-like inflation}},\
  }\href@noop {} {\bibfield  {journal} {\bibinfo  {journal} {Journal of High
  Energy Physics}\ }\textbf {\bibinfo {volume} {2023}},\ \bibinfo {pages} {1}
  (\bibinfo {year} {2023})}\BibitemShut {NoStop}%
\bibitem [{\citenamefont {Mirbabayi}\ and\ \citenamefont
  {Gruzinov}(2023)}]{Mirbabayi2023fudge}%
  \BibitemOpen
  \bibfield  {author} {\bibinfo {author} {\bibfnamefont {M.}~\bibnamefont
  {Mirbabayi}}\ and\ \bibinfo {author} {\bibfnamefont {A.}~\bibnamefont
  {Gruzinov}},\ }\bibfield  {title} {\bibinfo {title} {{Shapes of
  non-Gaussianity in warm inflation}},\ }\href@noop {} {\bibfield  {journal}
  {\bibinfo  {journal} {Journal of Cosmology and Astroparticle Physics}\
  }\textbf {\bibinfo {volume} {2023}}\bibinfo  {number} { (02)},\ \bibinfo
  {pages} {012}}\BibitemShut {NoStop}%
\bibitem [{\citenamefont {Behnel}\ \emph {et~al.}(2011)\citenamefont {Behnel},
  \citenamefont {Bradshaw}, \citenamefont {Citro}, \citenamefont {Dalcin},
  \citenamefont {Seljebotn},\ and\ \citenamefont {Smith}}]{Behnel2010cython}%
  \BibitemOpen
\bibfield  {number} {  }\bibfield  {author} {\bibinfo {author} {\bibfnamefont
  {S.}~\bibnamefont {Behnel}}, \bibinfo {author} {\bibfnamefont
  {R.}~\bibnamefont {Bradshaw}}, \bibinfo {author} {\bibfnamefont
  {C.}~\bibnamefont {Citro}}, \bibinfo {author} {\bibfnamefont
  {L.}~\bibnamefont {Dalcin}}, \bibinfo {author} {\bibfnamefont
  {D.}~\bibnamefont {Seljebotn}},\ and\ \bibinfo {author} {\bibfnamefont
  {K.}~\bibnamefont {Smith}},\ }\bibfield  {title} {\bibinfo {title} {Cython:
  The best of both worlds},\ }\href {https://doi.org/10.1109/MCSE.2010.118}
  {\bibfield  {journal} {\bibinfo  {journal} {Computing in Science
  Engineering}\ }\textbf {\bibinfo {volume} {13}},\ \bibinfo {pages} {31 }
  (\bibinfo {year} {2011})}\BibitemShut {NoStop}%
\end{thebibliography}%

\end{document}